\documentclass[prx,aps,amssymb,amsmath,reprint,superscriptaddress,floatfix,longbibliography]{revtex4-2}
\usepackage[utf8]{inputenc}
\usepackage{bm}
\usepackage{graphicx}
\usepackage{color}
\usepackage[b]{esvect} 
\usepackage{mathtools}
\usepackage{braket}
\usepackage{amsmath}
\usepackage{amsfonts}
\usepackage{amssymb}
\usepackage{textcomp} 
\usepackage[normalem]{ulem} 
\usepackage{hyperref}
\usepackage{float}
\usepackage{cases}
\usepackage{fancyhdr}

\newcommand{\pd}{{\phantom{\dagger}}}
\newcommand{\mr}{\mathrm}

\hypersetup{colorlinks=true}
\hypersetup{citecolor=red}

\begin{document}


\title{Simulating realistic screening clouds around quantum impurities: 
role of spatial anisotropy and disorder}

\author{Maxime Debertolis}
\affiliation{Institut N\'{e}el, CNRS and Universit\'e Grenoble Alpes, F-38042 Grenoble, France}
\author{Izak Snyman}
\affiliation{Mandelstam Institute for Theoretical Physics, School of Physics, University of the
Witwatersrand, Johannesburg, South Africa}
\author{Serge Florens}
\affiliation{Institut N\'{e}el, CNRS and Universit\'e Grenoble Alpes, F-38042 Grenoble, France}

\begin{abstract}
Dynamical quantum impurities in metals induce electronic correlations in real space that are difficult 
to simulate due to their multi-scale nature, so that only s-wave scattering in clean 
metallic hosts has been investigated so far. However, screening clouds should show anisotropy due to
lack of full rotational invariance in two- and three-dimensional lattices, while inherent
disorder will also induce spatial inhomogeneities. To tackle these challenges, 
we present an efficient and robust algorithm based on the recursive generation of natural 
orbitals defined as eigenvectors of the truncated single-particle density matrix. 
This method provides well-converged many-body wave functions on lattices with up to tens of thousands 
of sites, bypassing some limitations of other approaches.
The algorithm is put to the test by investigating the charge screening cloud around an
interacting resonant level, both on clean and disordered lattices, achieving accurate spatial
resolution from short to long distances.
We thus demonstrate strong anisotropy of spatial correlations around an adatom
in the half-filled square lattice.
Taking advantage of the efficiency of the algorithm, we further compute the disorder-induced
distribution of Kondo temperatures over several thousands of random realizations, at the 
same time gaining access to the full spatial profile of the screening cloud in each sample.
While the charge screening cloud is typically shortened due to the polarization of the impurity by
the disorder potential, we surprisingly find that rare disorder configurations preserve
the long range nature of Kondo correlations in the electronic bath.
\end{abstract}

\maketitle

\section{Introduction}
\label{Introduction}
The fundamental description of systems involving a macroscopic number of particles subject both to 
interactions and disorder is a central question in condensed matter physics~\cite{coleman_2015}. 
Quantum impurity problems, where a few localized degrees of freedom experience strong Coulomb 
repulsion while hybridizing with a much larger system of otherwise free particles,
constitute the simplest manifestations of many-body phenomena, which have made them a central focus 
of research~\cite{CostiRMP}. Beyond their obvious relevance 
in describing actual impurities in metals~\cite{Hewson_Book} or transport in electronic quantum 
dots~\cite{Costi_Transport},
quantum impurity models are used in a wider context through the dynamical mean field theory
(DMFT)~\cite{Georges_Kotliar_Krauth_Rozenberg_1996}, describing at the local level, fully interacting
lattice problems.
Over the years, considerable effort has been devoted to the study of spatial correlations between a quantum 
impurity and its surrounding environment. 
[See Ref.~\onlinecite{Affleck_Review} for a review.]
The associated ``Kondo cloud" indeed reveals a non-trivial screening process extending deep into
the Fermi gas~\cite{Mitchell_Cloud}, that is however challenging to access experimentally through transport measurements.
Significant theoretical attention has focused on 
the subtle correlations displayed in the spatial profile of the 
electronic density in response to a local polarization of the
impurity~\cite{Gubernatis_1987,Sorensen_Scaling_1996,Barzykin_Affleck_1998,Borda_2007,Holzner_Cloud_DMRG,
Dagotto_Cloud,Florens_Snyman_Universal}. 
Also, many alternative proposals for measurements have been 
made~\cite{Simon_Cloud,Cornaglia_Cloud_FiniteSize,Monien_Cloud_FiniteSize,Halperin_Cloud_Staircase,
Katsnelson_Tunneling_2009,Sim_Cloud_Proposal,
Nishida_Cloud_ColdAtoms,Bauer_Cloud_ColdAtoms,Snyman_Cloud_CQED,Zarand_Cloud_SC,Cohen_Dynamics_2021}, 
and recently experimental evidence for a characteristic Kondo length was obtained~\cite{Borzenets2020}.


Inherent to the experimental realizations of dynamical quantum impurity systems is the
presence of charged or magnetized static defects that can affect the properties of the bulk metallic host.
Strong randomness in an environment harboring dilute magnetic impurities can for instance induce 
wide distributions of Kondo
temperatures~\cite{Dobrosavljevic_Kirkpatrick_Kotliar_1992, Kettemann_Mucciolo_2006}, 
leading to exotic non-Fermi liquid behavior.
In the standard case of the spin Kondo effect, it was demonstrated that the random potential in
the bath decreases the density of states at the impurity
site~\cite{Cornaglia_Grempel_Balseiro_2006},
hence exponentially suppressing the Kondo temperature, which makes, paradoxically, the Kondo
screening cloud extend over much longer scales than in the clean case. Clearly, a microscopic picture
of the spatial Kondo correlations at play in dirty metals is still missing. 
In addition, studies of screening clouds in higher than one dimension have relied until now on full
circular/spherical symmetry of the Fermi surface, reducing the problem to one-dimension in the
case of s-wave impurity scattering. However, more complicated Fermi surfaces can give rise to anisotropic 
screening clouds, a topic that has also not been investigated yet. Our main goal here is to
provide a tailored quantum impurity solver that can efficiently resolve spatial correlations on
large-scale lattices, and possibly in the presence of spatial disorder, allowing us to explore some 
surprising aspects of these systems.

Simulating the real space screening properties of disordered quantum impurity problems requires 
not only techniques that can tackle lattices with tens of thousands of sites, but that are also 
fast enough to systematically sample over thousands of disorder configurations. Indeed, Numerical 
Renormalization Group
(NRG)~\cite{Krishna-murthy_Wilkins_Wilson_1980,CostiRMP} studies of Kondo
clouds~\cite{Borzenets2020} have been limited so far to clean
environments~\cite{Borda_2007,Lechtenberg},
due to the time-consuming reconstruction of accurate
real space features.
The Density Matrix Renormalization Group (DMRG)~\cite{White_1992} can also
resolve real space features around quantum impurities~\cite{Holzner_Cloud_DMRG}, but 
existing studies are limited to lattices with a few hundreds of sites only.
As a result, only approximate scaling equations have been used to provide a (qualitative) picture 
of how the single Kondo scale is affected by disorder~\cite{Cornaglia_Grempel_Balseiro_2006}.
Understanding quantitatively the microscopic properties of disordered quantum impurity problems 
and the associated spatial correlations in the dirty screening cloud remains an open problem, that we 
wish to address here.


In order to tackle the challenge of simulating dirty quantum impurity problems on large
lattices and for a huge number of disorder samples, we propose to exploit the hierarchical structure 
of electronic correlations when impurity problems are expressed in their
\textit{natural orbital} (NO) basis. In terms of second quantization fermionic 
operators $c_{i}^\dag$ describing an arbitrary complete set of orbitals,
natural orbitals $q^\dagger_\alpha=\sum_j c_j^\dagger P_{j\alpha} $ are constructed from the orthonormal 
eigenvectors $P_{j\alpha}$ obeying $\sum_j Q_{ij}P_{j\alpha}=\lambda_\alpha P_{i\alpha}$, with
$Q_{ij}$ the one-body density matrix (covariance matrix) defined
as $Q_{ij}=\langle c^{\dagger}_{i}c^{\pd}_{j}\rangle$, where the expectation value is 
with respect to the full 
many-body ground state of the problem.
The efficient representation of many-body problems through NOs was first exploited in 
quantum chemistry
\cite{Lowdin_1955,Davidson_1972}, permitting an optimal description of molecular wave functions,
while taking into account the most relevant correlations at play. Indeed, core orbitals of
atoms or molecules are almost fully occupied, and correspond to frozen NOs with an occupation close
to 2 (considering spin-1/2 electrons), that can be well-treated at 
mean field level. Valence orbitals constitute
the active space, that must be attacked with a full configuration interaction calculation, which
allows one to deal with a smaller and most pertinent subspace of the complete Hilbert space.

However, NOs are not known beforehand, as they depend on the full ground state of the system, 
and numerical methods have been developed to approximate them.
Typically, the active space is constructed by advanced minimization algorithms,
that form the core of standard quantum
chemistry packages~\cite{CASSCF_Review}. Iterative schemes have also been proposed in the quantum chemistry
context by Li and Paldus~\cite{Li_Paldus_2005}, and the recursive algorithm
that we present in this article exploits similar ideas, yet in the field of quantum impurity
models. Indeed, it has recently been demonstrated numerically that the ground states of quantum impurity models 
in clean hosts are efficiently described by
NOs~\cite{He_Lu_2014,Yang_Feiguin_2017,Debertolis_Florens_Snyman_2021}, and a rigorous mathematical
background for these ideas has also been provided~\cite{Bravyi_Gosset_2017}.
These studies of the covariance matrix in impurity models have shown that approximations in which
correlations are confined to a small subset of NOs, converge exponentially to the true ground state
as the number of correlated orbitals is increased ~\cite{Debertolis_Florens_Snyman_2021}. 

This observation allows two alternative yet efficient descriptions of the ground state of quantum
impurity models.
The first description relies on trial wave functions that are linear superpositions of a few non-orthogonal Gaussian
states~\cite{Bravyi_Gosset_2017,Boutin_Bauer_2021,Snyman_Florens_2021}. 
This method relies on optimizing the trial state over the manifold of
parameters defining the Slater determinants, a task that becomes harder the larger the system.
The second description uses a strict separation of active (correlated) and
inactive (uncorrelated) orbitals in the one-particle natural basis, treating the former in
full configuration interaction, and the latter within a single Slater
determinant~\cite{Li_Paldus_2005,Debertolis_Florens_Snyman_2021}. This is the approach that we will
pursue in the present work, although extending the Gaussian state ideas might be fruitful as well.

In the context of quantum impurity problems, and in some cases within DMFT calculations, various
algorithms based on natural orbitals have already
been developed~\cite{Zgid_Gull_Chan_2012, 
Lu_Hoppner_Gunnarsson_Haverkort_2014, Lu_Cao_Hansmann_Haverkort_2019, Kitatani_Sakai_Arita_2021},
some of these methods being approximate, while others allow for systematic 
improvement and can reach a ground state error that is
orders of magnitude smaller than the smallest relevant energy scale for model parameters 
of practical interest. We employ a method~\cite{Li_Paldus_2005, Lin_Demkov_2013, He_Lu_2014} that 
recursively generates improved guesses 
for the NO basis by incorporating new orbitals into the active space and discarding old ones
in steps reminiscent of sweeps in the Density Matrix Renormalization Group. 
By demonstrating the usefulness of iterative solvers in the dirty case,
we extend the scope of methods based on natural orbitals, especially for a more realistic
description of interacting and disordered systems, a notoriously difficult problem.
We have experimented with various sweeping protocols, and present one here that 
leads to rapid convergence in most realizations of the disordered interacting resonant level model
that we studied. We believe the same protocol will prove efficient for other disordered impurity models.

Our results demonstrate a wealth of counter-intuitive results that we briefly summarize here.
For a half-filled square lattice, correlations are always negative in the non-interacting case, but
interactions can drive positive correlations between the impurity and its host not only at short distances, 
but on a ``butterfly pattern'' at a finite range. These correlations die off mostly faster than with 
s-wave symmetry, but spread to longer distances on the diagonal of the 2D lattice. 
In the case where the metallic host is disordered, we also find surprisingly that some rare configurations 
can survive the presence of strong disorder, with screening clouds that are robust thanks to
orbitals living near the Fermi level.
These two observations are relevant for the understanding of the entanglement generated by a Fermi liquid
between diluted quantum impurities in realistic lattices. 

The article is structured as follows. In Sec. \ref{RGNO}, we present the recursive algorithm for 
quantum impurity models, together
with the sweeping protocol that, in our experience, gives the best performance. 
We then benchmark the method on the
Interacting Resonant Level Model (IRLM), by comparing 
to accurate NRG simulations on the Wilson
chain. We also present the direct computation of the Kondo screening cloud in the clean case, for a large
number of sites in a real space chain, a method that is more practical than NRG simulations. 
Having demonstrated the method's ability to handle large systems, we then employ it to study 
the highly anisotropic screening cloud that results in a two-dimensional square lattice at half-filling.
In Sec.~\ref{Disorder}, we use the recursive algorithm to study the effect of charge disorder on the bath 
surrounding a quantum impurity, and give an analysis of spatial correlations within the dirty screening cloud.
Finally, in Sec.~\ref{Conclusion}, we summarize our results and present potential applications for
disordered Kondo systems. Two appendices discuss technical matters.

\section{Recursive Generation of Natural Orbitals}\label{RGNO}

\subsection{The general idea}

Natural orbital methods are based on a many-fermion variational ground state Ansatz of the form
\begin{equation}
\left|\Psi\right>=f_1^\dagger\ldots f_{N_{\rm occ}}^\dagger \sum_S\Psi_S\left|S\right>.
\label{ansatz}
\end{equation}
Here $f_1^\dagger$ to $f_{N_{\rm occ}}^\dagger$
create fermions in orbitals deep enough in the Fermi see
that their occupancy can be approximated as equal to one. We refer to this set as the occupied sector.
From the orthogonal complement of the occupied sector, one defines a reduced set of $M$ orthogonal orbitals
that will be hosting the correlations. 
We call this subspace the correlated sector (or active space in quantum chemistry).
The remaining orbitals define the unoccupied sector, and, with an occupancy
approximated to zero, do not take part in the state~(\ref{ansatz}).
The sum over $S$ in (\ref{ansatz}) runs over all Slater determinants
describing a certain number $N_\mathrm{cor}$ of fermions occupying orbitals in the correlated sector
(typically $M/2$ particles in case of half-filling).
For a given choice of the 
occupied, correlated and unoccupied sectors, the coefficients $\Psi_S$ that minimize
the expectation value of the energy, are such that $\sum_S\Psi_S\left|S\right>$ is the ground state
of an effective Hamiltonian $H_{\rm eff}$ in the correlated sector, which is obtained from
the full Hamiltonian $H$ by treating particles in the occupied sector at mean field level. (Full details
are provided below.) The quality of the Ansatz hinges on the choice of orbitals that span the 
occupied, correlated and unoccupied sectors. The optimal choice has the property that
the occupied and unoccupied sectors are spanned by the eigenvectors (natural orbitals) 
of the covariance matrix that
have eigenvalues closest to one or zero respectively. In fermionic impurity problems,
the energy expectation value of the optimal Ansatz converges exponentially to the true ground state energy
as $M$ is increased, with a rate that stays finite in the thermodynamic limit.

The true covariance matrix of the system is not known a priori, as it needs to be calculated from the
exact ground state. Strategies that recursively generate improved guesses for the NOs
surmount this problem as follows. Given is a guess for the NOs and a partitioning into 
the occupied, correlated and unoccupied sectors. To improve this guess, the correlated
sector is expanded by adding to it well-chosen orbitals, typically one each from the current 
occupied and unoccupied sectors, as well as increasing the number of particles hosted
by the correlated sector, typically by one. This results in a correlated state with typically 
one added fermion, that we denote here
$\sum_S\Psi^\mr{add}_S\left|S\right>$, obtained as the ground state of the effective 
Hamiltonian $H_{\rm eff}$ in the enlarged correlated sector.
From there, the covariance matrix is estimated as 
$Q_{ij}=\left<\Psi^\mr{add}\right|c_i^\dagger c_j\left|\Psi^\mr{add}\right>$,
where $\left|\Psi^\mr{add}\right>=f_1^\dagger\ldots f_{N_{\rm occ}-1}^\dagger
\sum_S\Psi^\mr{add}_S\left|S\right>$ is the full many-body state obtained by completing 
$\sum_S\Psi^\mr{add}_S\left|S\right>$ with the $N_\mr{occ}-1$ fermions in the reduced occupied sector.
This estimate for the covariance matrix has $M+2$ eigenvectors associated with eigenvalues
different from zero and one. Of this set, the two eigenvectors with eigenvalues closest to zero and one
are then respectively put back into the occupied and unoccupied sectors, thus
reducing the number of orbitals in the correlated sector back to $M$. At this point the procedure is 
repeated, starting with the step of choosing orbitals from the occupied and unoccupied sectors
to add to the correlated sector.
The protocol for making this choice is designed to
ensure that all (or most) steps lower the energy. The process is repeated until $\left|\Psi\right>$
has flowed sufficiently close to a fixed point. In general, this fixed point may not be completely optimal.
However, our benchmarking of the protocol we devised against numerically exact NRG results
reveals that the error of the fixed point of our protocol, compared to the optimal state of form (\ref{ansatz})
is negligible with respect to the deviations to the exact ground state.

\subsection{The detailed algorithm}

\begin{figure}[ht]
\centering
\includegraphics[width=0.99\columnwidth]{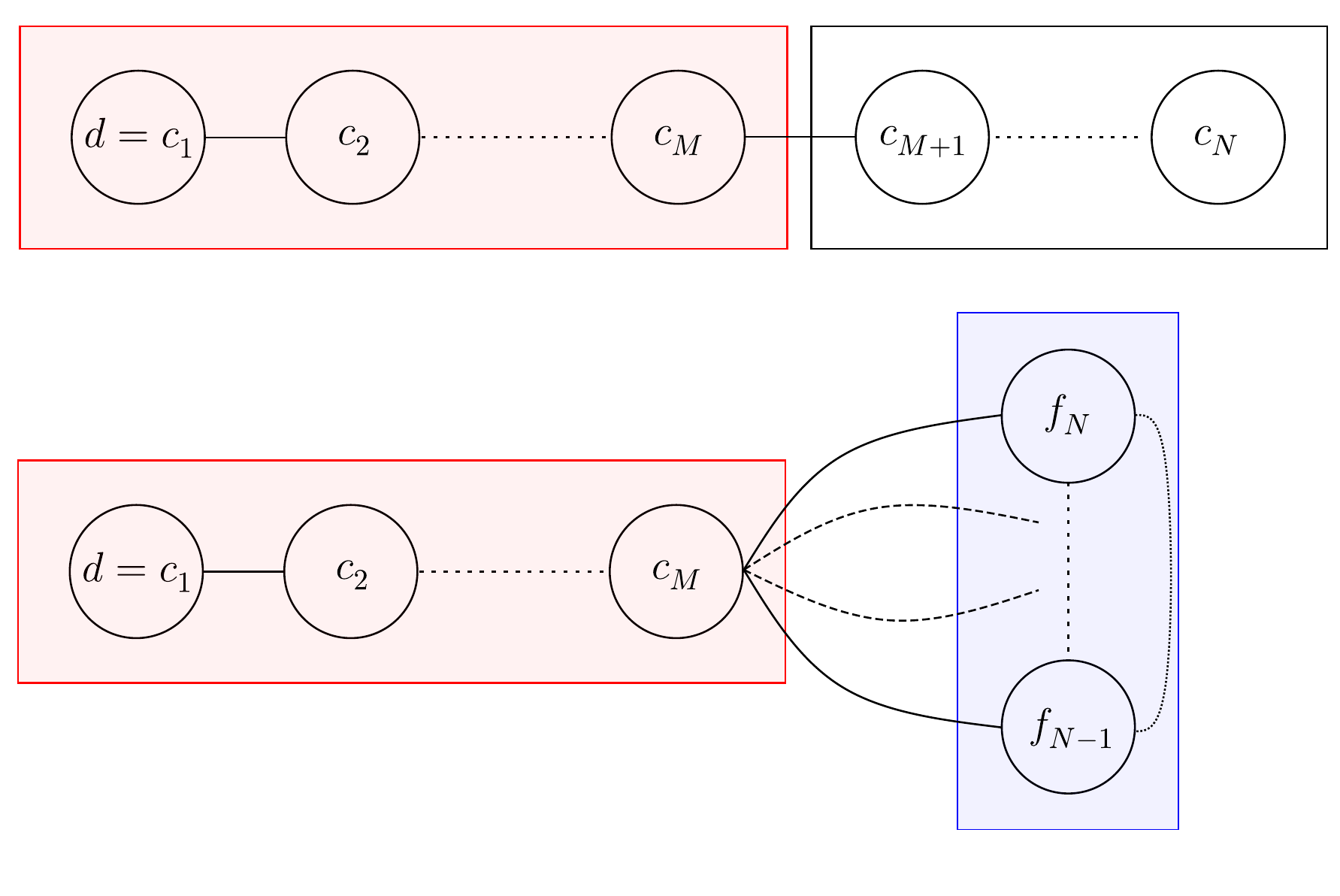}
\caption{Initialization:
The first $M$ sites of the chain are chosen to describe
the correlated space (in red), whereas a diagonalization of the hopping matrix
$\hat{t}=\hat{P}\hat{E}\,^{t}\hspace{-0.2em}\hat{P}$ between the remaining $N-M$ sites defines the
orbitals spanning the uncorrelated space (in blue). Orbitals $f^{\protect\pd}_{\alpha}$ 
obtained from this diagonalization are paired according
to their energy relative to the chemical potential, and introduced into
the correlated sector two at a time, before proceeding to the diagonalization of the enlarged
space.}
\label{init_c}
\end{figure}

The efficiency of the recursive generation of natural orbitals crucially depends on the initial 
choice of the NOs and 
the recursive update protocol. We now present the strategy that we found most reliable and efficient.
Our protocol is generic, but it is convenient to introduce a concrete model in order to 
explain it. Consider therefore the clean Interacting Resonant Level Model, whose disordered
version we will study in detail in Sec.~\ref{Disorder}:
\begin{eqnarray}
\nonumber
H &=& V\left(c_1^{\dagger}c^{\pd}_{2} + {\mathrm h.c.}\right) +
\sum\limits_{i=2}^{N-1} t^{\pd}_{i}\left(c^{\dagger}_{i}c^{\pd}_{i+1} + \mathrm{h.c.}\right) \\ 
&&+ U\left(c_1^{\dagger}c_1 -\frac{1}{2} \right) \left( c^{\dagger}_{2}
c^{\pd}_{2} - \frac{1}{2}\right).
\label{IRLM}
\end{eqnarray}
Here $c^{\dagger}_{i}$ creates an electron at site $i$ of the chain, with $N$ sites in
total. See top panel of Fig.~\ref{init_c}. 
This spinless model consists of a fermionic impurity (at site 1)
coupled to a non-interacting electronic lead, represented by a tight-binding chain. Despite 
the spinless nature of Eq.~(\ref{IRLM}), Kondo physics 
emerges for $U<0$ in the charge sector (here $U$ is the Coulomb interaction between the impurity
and the bath).
This is due to the degeneracy of the states $|0\rangle$ and
$ c_1^\dag c_2^\dag |0\rangle$ that is not lifted at first order in $V$, the hybridization
between impurity and the bath.
A chain form of the bath is used for simplicity, but the
method presented here can be adapted to any lattice, even in 2D or 3D. (See Appendix~\ref{appA}).
The values of the hopping parameters $t^{\pd}_{i}$
in the chain are for now left arbitrary, and are restricted to nearest neighbors here just for simplicity
of notation. 

\textbf{Initial choice for the correlated sector.} We first have to choose a complete set of orbitals in order to start the
iteration procedure. In this initial basis, we also need to pick $M$ orbitals that will constitute the
correlated space.
Convergence will clearly be helped if the initial guess is closer to the final solution.
Bearing in mind that we eventually will study disordered systems, 
we choose to initialize the correlated space by selecting
the impurity orbital and the Wannier orbitals associated with the $M-1$ sites nearest to the
impurity 

\textbf{Iterative diagonalization.}
For a given choice of the correlated sector, the occupied and unoccupied sectors must be constructed
from orbitals that span the orthogonal complement of the correlated sector. We note that an arbitrary 
rotation and re-ordering of these orbitals is possible since they are nearly degenerate with
respect to the spectrum of the covariance matrix, and we found that such an update can strongly affect 
the convergence speed.
During the iterative procedure of building the natural orbitals, the orbitals initially defined as uncorrelated 
will be added to the correlated space two at a time, and it is important 
to choose the most favorable set. We want to start with those that are most likely to
participate in correlations, which are generally the ones closest to the Fermi energy.
We thus perform a one-body diagonalization
of the non-interacting part of the Hamiltonian, projected onto the orthogonal complement of the
correlated sector:
\begin{equation}
\sum_{j=1}^N\left<0\right|c_i \mathcal P \left.H\right|_{U=0}\mathcal P c_j^\dagger\left|0\right>P_{j \alpha}=
E_\alpha P_{i\alpha}
\end{equation}
where $\mathcal P=\sum_{\alpha}f^\dagger_\alpha\left|0\right>\left<0\right|f_\alpha$
and $\left\{f^\dagger_\alpha\right\}$ spans the orthogonal
complement of the current correlated sector. 
In the first step, we can choose 
$\{f^\dagger_\alpha\}=\{c_j^\dagger|j=M+1,\ldots,N\}$, as
illustrated in Fig.~\ref{init_c}.
We use the results of this diagonalization to update the bases for the occupied and unoccupied
sectors
$f^{\dagger}_{\alpha} \leftarrow \sum_{i=1}^{N}P_{i\alpha} c^\dagger_{i}$,
such that orbitals $f^{\dagger}_{\alpha}$ with $E^{\pd}_{\alpha}<0$ 
form the occupied sector and the remaining ones form the
empty sector.
We sort the uncorrelated orbitals 
such that $E_1$ is the first energy below the Fermi energy, $E_2$ is the first energy above the
Fermi energy, $E_3$ is the second energy below the Fermi energy, etc.,
and use this order when we incorporate orbitals into the correlated
sector during the iterative process of building the natural orbitals.

After the above determination of the correlated and uncorrelated sectors, we apply the following
sweep protocol. 
We denote the current 
 set of correlated orbitals $q^{\dagger}_1,\ldots,q^{\dagger}_M$. 
Their expansion coefficients in terms of the site basis $c^{\dagger}_i$ are denoted $P_{ia}$. 
(At step $n=1$,
the $q_a^\dagger$ orbitals are equal to the $M$ first site orbitals $c_i^\dagger$.) In the first
step of the sweep, we remove the first pair of uncorrelated orbitals $f_1^\dagger$ and $f_2^\dagger$
from the occupied and unoccupied sectors respectively and add them to
the correlated sector : $q^{\dagger}_{M+1}= f^{\dagger}_1$ and $q^{\dagger}_{M+2}=f^{\dagger}_2$.
All other $f$-orbitals are kept frozen,
resulting in a few-body Hamiltonian truncated to the $M+2$ kept orbitals (see
Ref.~\onlinecite{Debertolis_Florens_Snyman_2021} for details):
\begin{eqnarray}
\label{Hfew}
H_{\rm eff} &=&\sum\limits_{a,b=1}^{M+2} \left[t^{\pd}_{ab}+2\sum\limits_{\alpha=3}^{N-M}
\left(U_{\alpha a b\alpha} - U_{\alpha a\alpha b}\right)n_\alpha \right]
q^{\dagger}_{a} q^{\pd}_{b}\nonumber\\
&+& \sum\limits_{a,b,c,d}U^{\pd}_{abcd}\;
q^{\dagger}_{a}q^{\dagger}_{b}q^{\pd}_{c}q^{\pd}_{d}+E_{\rm occ},\\
E_{\rm occ}&=& \sum\limits_{\alpha=3}^{N-M}\!\!\!t^{\pd}_{\alpha\alpha}n^{\pd}_\alpha
+ \sum\limits_{\alpha,\beta=3}^{N-M}(U^{\pd}_{\alpha\beta\beta\alpha}-U^{\pd}_{\alpha\beta\alpha\beta})
n^{\pd}_\alpha n^{\pd}_\beta.
\label{Eocc}
\end{eqnarray}
Here the latin indices $a,b,c,d$ refer to the $M+2$ correlated orbitals, while the greek
indices $\alpha,\beta$ denote the $N-M-2$ frozen orbitals $f^{\dagger}_\alpha$ that have
occupancy $n^{\pd}_\alpha=0$ or 1, providing the mean-field shift of the 
 first line of Eq.~(\ref{Hfew}).
To obtain this expression, we reformulated the single interaction term $U c^\dagger_1 
c^\dagger_2 c^\pd_2 c^\pd_1$ term of Hamiltonian~(\ref{IRLM}) as a four-leg tensor
$U_{ABCD} = U P_{1A} P_{2B} P_{2C}^* P_{1D}^*$ in the complete NO basis 
 i.e. $A,\,B,\,C$ and $D$ can each be a lower case roman index referring 
to a correlated orbital or a greek index referring to an uncorrelated orbital.
As a result, $U_{abcd}$ is the contribution that is internal to the correlated $q$-orbitals, while
$U_{\alpha\beta\gamma\delta}$
acts within the uncorrelated sector, and provides only a constant energy shift in~(\ref{Eocc}).
Terms like $U_{a\alpha b \alpha}$ couple both sectors without exchanging particles and 
contribute to the additive one-body part of the correlated sector.
Similarly, 
\begin{eqnarray}
t_{AB}&=&-\frac{U}{2}(P_{1A}P_{1B}^*+P_{2A}P_{2B}^*)+V(P_{1A}P_{2B}^*+P_{2A}P_{1B}^*)\nonumber\\
&+&\sum_{i=2}^{N-1} t_i (P_{iA}P_{i+1,B}^*+P_{i+1,A}P_{iB}^*),
\end{eqnarray}
 describes all the additive one-body terms of
$H$ in the NO basis, accounting
for the chain hoppings $t_i$, the hybridization $V$, and the particle-hole
symmetry restoring potential $-U(n_1+n_2)$ in~(\ref{IRLM}).
Only the terms $t_{ab}$ and $t_{\alpha\alpha}$ contribute in~(\ref{Hfew}), because
correlated and uncorrelated orbitals do not mix in the wave function.
The choice 
$\sum_S\Psi^\mr{add}_S\left|S\right>=\left|{\rm GS}\rm\right>$ where $\left|{\rm GS}\rm\right>$
is the $N_{\rm cor}+1$ particle ground state of $H_{\rm eff}$ (with one particle added) minimizes 
with respect to the coefficients $\Psi^\mr{add}_S$ the 
expectation value $\left<H\right>$ of the full Hamiltonian
for the complete Ansatz 
$\left|\Psi^\mr{add}\right>=f_{1}^\dagger\ldots f_{N_{\rm occ}-1}^\dagger \sum_S\Psi^\mr{add}_S\left|S\right>$.
The matrix dimension of $H_{\rm eff}$ grows exponentially in $M$,
but because the original Hamiltonian typically only contains two-body interactions, the number
of non-zero entries per row only grows like $M^4$. We could therefore efficiently find the 
ground state of $H_{\rm eff}$ using sparse matrix techniques for $M\leq 16$, although 
$M=6$ already proved sufficiently accurate for most practical purposes.

\begin{figure}[h!]
\includegraphics[width=0.99\columnwidth]{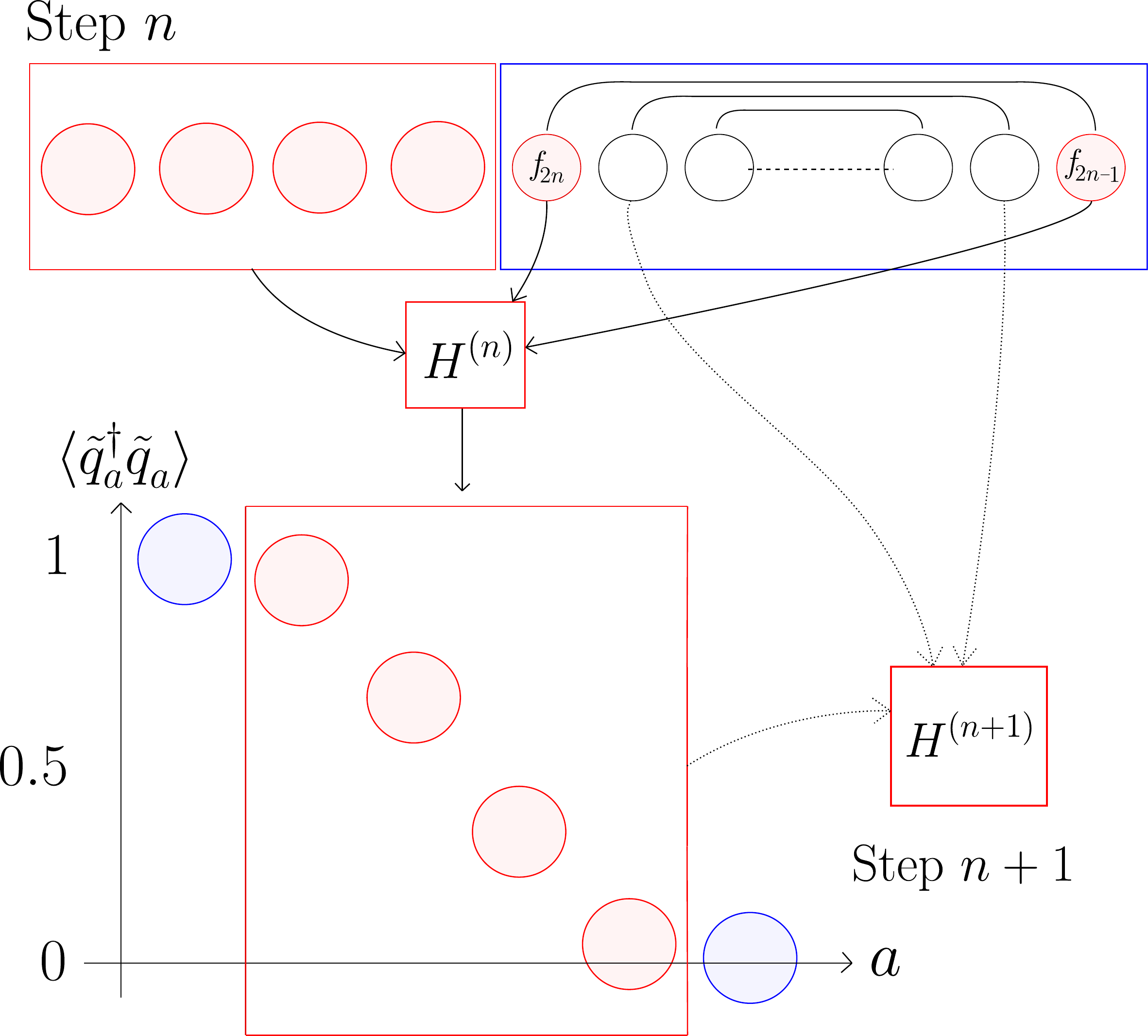}
\caption{Evolution of the Hamiltonian during step $n+1$
of a sweep. We define $H^{(n)}$ by adding
the orbitals $f^{\protect\pd}_{2n-1}$ and $f^{\protect\pd}_{2n}$ to the current correlated space. Computing
the covariance matrix $\hat{Q}$ in the ground state of $H_{n}$ defines a new set of orbitals
$\tilde q^{\protect\pd}_i$.
Dropping the two orbitals (in blue) that have occupancy closest to 0 and 1, we can add two new
$f$-orbitals, thus defining the Hamiltonian $H^{(n+1)}$ at step $n+1$.}
\label{End_sweep}
\end{figure}

From the ground state of $H_{\rm eff}$, the non-trivial $(M+2)\times(M+2)$ block 
$Q_{ab}=\langle q^\dagger_a q^\pd_b\rangle$ of the covariance matrix is obtained,
and its eigenvectors provide a
new set of $M+2$ orbitals $\tilde q^{\dagger}_a$.
The two $\tilde q$-orbitals with covariance matrix
eigenvalues closest respectively to $0$ and $1$ are
redefined as new $f$-orbitals and moved to the unoccupied and occupied spaces.
The $M$ kept orbitals define the new correlated space, to which the next two $f$-orbitals are added.
This diagonalization procedure is iterated until all uncorrelated $f$-orbitals have been
incorporated into the correlated sector once.
See Fig.~\ref{End_sweep} for a schematic description of this whole scheme. This process
allows inclusion of information from all orbitals (describing all spatial scales) into the update of the
correlated orbitals.

\textbf{Sweeps.} The iterative diagonalization described above constitutes a sweep, that gives a
first approximation of the NOs of the problem. In order to reach full convergence of the NOs
(for a given size $M$ of the correlated subspace), we perform several sweeps, taking the NOs obtained
at the end of the previous sweep to initialize the next one. Monitoring the energy or any other
relevant observable provides some practical notion of convergence of the NOs, and in most cases, 10
to 20 sweeps are enough. The accuracy of the final ground state can generally be assessed by computing
the variance of the total Hamiltonian, but we turn now to a more detailed comparison to NRG results.

\subsection{Benchmark on the Wilson chain}
\label{benchmark}
We test the quality of the many-body wave function our iterative algorithm produces by 
comparing to NRG, taking the same energy grid in the bath for both algorithms.
We model the bath with a constant density of states in a symmetric band of half-width $D$, that we take as
our unit of energy. We perform the standard preliminaries of logarithmic discretization
of the bath energies, 
transforming the bath Hamiltonian into a Wilson chain with hopping parameters:
\begin{equation}
t_{i}=\frac{\left(1+\Lambda^{-1} \right)\left(1 -
\Lambda^{-(i-2)-1}\right)}{2\sqrt{1-\Lambda^{-2(i-2)-1}}\sqrt{1- \Lambda^{-2(i-2)-3}}}
\Lambda^{-(i-2)/2}D.
\end{equation}
Here the index $i$ has been offset as $i-2$ with respect to the usual NRG conventions, because the chain
starts at $i=2$ and not $i=0$ in our
definition for Hamiltonian~(\ref{IRLM}). We perform simulations for $\Lambda=2$, $N=110$ sites and
$N_{\mathrm{kept}}=1500$ kept states, which ensures converged results close to the thermodynamic limit 
given the parameters of the following study.

When the chemical potential of the bath is set to $\mu=0$, Hamiltonian~(\ref{IRLM}) enjoys
particle-hole symmetry, and the ground state exhibits a quantum phase transition for 
 $U\simeq-1.3D$,
and well-developed Kondo correlations for $U<0$, which permits us to 
 test our methodology in a nontrivial many-body regime.
Nevertheless, we have to be careful with both algorithms when we approach close to the transition,
since the characteristic energy scale $T_{\mathrm{K}}$ vanishes exponentially in this regime,
 whereas the thermodynamic limit requires $N> D/T_\mathrm{K}$.
For the IRLM, where Kondo correlations develop in the charge sector, the Kondo temperature is
defined as:
\begin{equation}
T_{\mathrm{K}} = \frac{1}{4\chi}, \;\; \mathrm{with} \;\; \chi = \frac{\partial\langle \hat{n}^{\pd}_{1}
\rangle}{\partial \epsilon^{}_{1}}\Big|_{\epsilon^{}_{1} \rightarrow 0}
\label{Tk}
\end{equation}
where $\chi$ is the local susceptibility of the impurity to a small energy bias 
$\epsilon^{\pd}_1 \hat{n}^{\pd}_{1}$, with $\hat{n}^{\pd}_{1}=c^{\dagger}_{1}c^{\pd}_{1}$ 
the density operator at the impurity site $i=1$.
The Kondo temperature 
measures the stability of the ground state to local particle-hole symmetry breaking perturbations 
at the site
of the impurity, and its vanishing at the critical point signals the onset of spontaneous particle-hole
symmetry breaking at $U<U_\mathrm{c}$. 
In practice, the derivative is computed by 
 calculating the ground state twice, once at $\epsilon_1=0$ and 
 once at $\epsilon_1\ll T_\mathrm{K}$.
\begin{figure}[ht!]
\includegraphics[width=1\columnwidth]{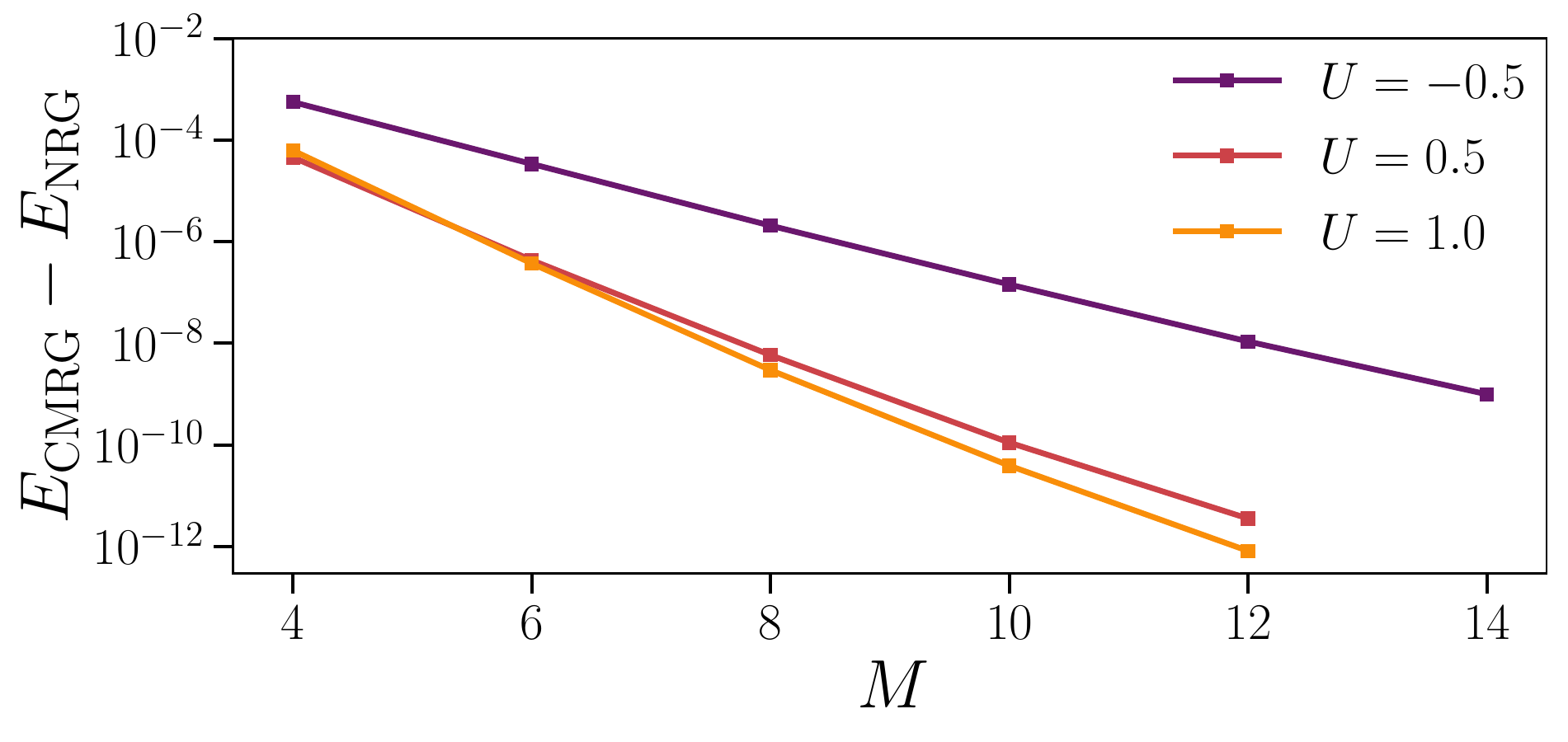}
\caption{IRLM ground state energy obtained from the recursive natural orbitals algorithm,
relative to the NRG result, as a function
of the number of correlated orbitals $M$, and for three different values of the interaction $U$.
The calculation is performed for hybridization $V=0.15D$, Wilson parameter $\Lambda=2.0$, and
$N=110$ sites.}
\label{Conv_M}
\end{figure}
In Fig.~\ref{Conv_M}, we compare the ground state energy obtained by 
 recursive generation of natural orbitals
to the one computed
in NRG, for different values of the interaction, and hence for different $T_{\mathrm{K}}$.
We see that the natural orbital
algorithm converges exponentially with the number $M$ of natural orbitals in
the correlated space for any value of the interaction as previously reported
\cite{Bravyi_Gosset_2017,Debertolis_Florens_Snyman_2021}, thanks to the exponential decay of
the occupancies in the covariance matrix of quantum impurity models.
In contrast to the NRG-based determination of the natural orbitals in
Ref.~\onlinecite{Debertolis_Florens_Snyman_2021}, the correlated orbitals are
now directly found through the recursive protocol described in Sec.~\ref{RGNO}.
For $U=-0.5D$, the ground state energy is converged to 5 significant digits for only $M=6$ correlated orbitals,
and we can go up to 8 digits with $M=12$. This precision is well below the corresponding value
of $T_{\mathrm{K}} \simeq 3.10^{-3}D$, ensuring that Kondo correlations are fully captured.
In weakly correlated regimes, either for $U>0$ or for $U\ll-1.3D$, we can easily converge the
energy with a precision better than $10^{-10}$. The non-interacting case $U=0$ is
exact by construction of the ansatz~(\ref{ansatz}).
\begin{figure}[h!]
\includegraphics[width=1\columnwidth]{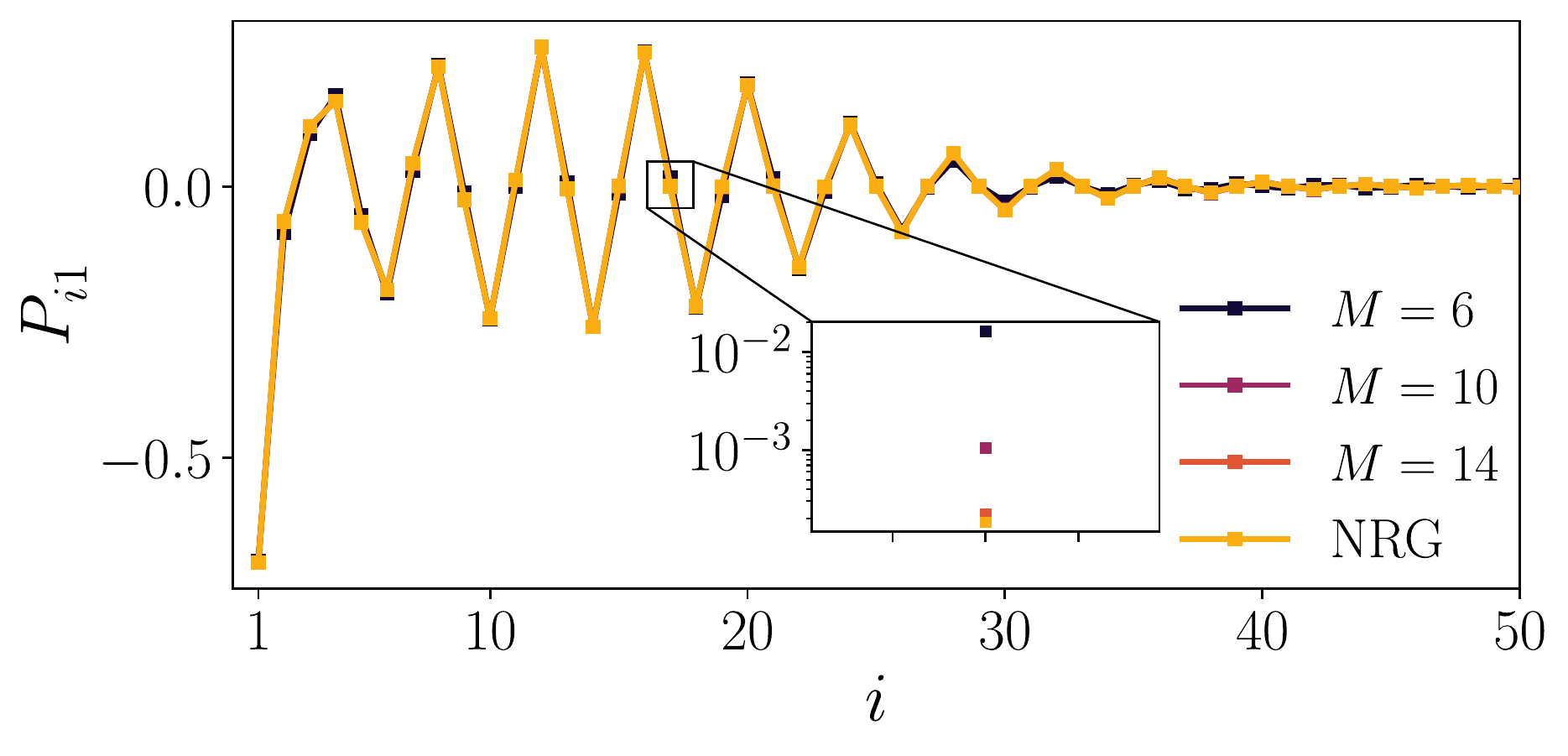}
\caption{Amplitude $P_{i1}$ along the Wilson chain of the most correlated natural orbital $q^{\protect\pd}_1$,
computed in NRG and with the recursive natural orbitals algorithm, for $U=-0.5D$. 
This orbital is the eigenvector of the $\hat{Q}$ matrix whose
eigenvalue is the closest to $1/2$. The inset shows the convergence with $M$ at site $n=16$,
and similar precision is obtained for all sites. The simulation is done for a chain of
$N=110$ sites, but the plot is cut at $N=50$ since the weight of the orbital is almost $0$
further on.}
\label{Conv_NO1}
\end{figure}

Another quantity of interest is the ``spatial'' extension of the NOs along the Wilson
chain. The complete set of natural orbitals can be extracted from converged NRG
simulations~\cite{Debertolis_Florens_Snyman_2021} by computing the full covariance matrix
$\hat{Q}$, and extracting its eigenvectors. We compare in Fig.~\ref{Conv_NO1} the 
 NRG and recursively obtained
amplitude $P_{i1}$ of the most correlated orbital $q^{\dagger}_1=\sum_i P^{\pd}_{i1} c^{\dagger}_i$ 
(defined as having
its occupancy the closest to 1/2), which carries most of the Kondo entanglement. Even with a number
of correlated orbitals as low as $M=6$, the recursive result
is nearly indistinguishable from the
NRG results (the inset shows a precision of several digits, that improves by increasing $M$).
We see in Fig.~\ref{Conv_NO1} that the largest amplitudes are around the
site $n=15$, which corresponds to an energy of $\Lambda^{-15/2}D \simeq 5.10^{-3}$
comparable indeed to $T_{\mathrm{K}}\simeq 3.10^{-3}$.
At fixed $M$, the recursive algorithm can only find the $M$ most 
correlated orbitals, but this is not a real disadvantage in comparison to other methods: 
due to their very near degeneracy in the spectrum of $\hat{Q}$, alternative methods also
have to apply exponential effort to resolve the remaining orbitals individually.

\begin{figure}[h!]
\includegraphics[width=1\columnwidth]{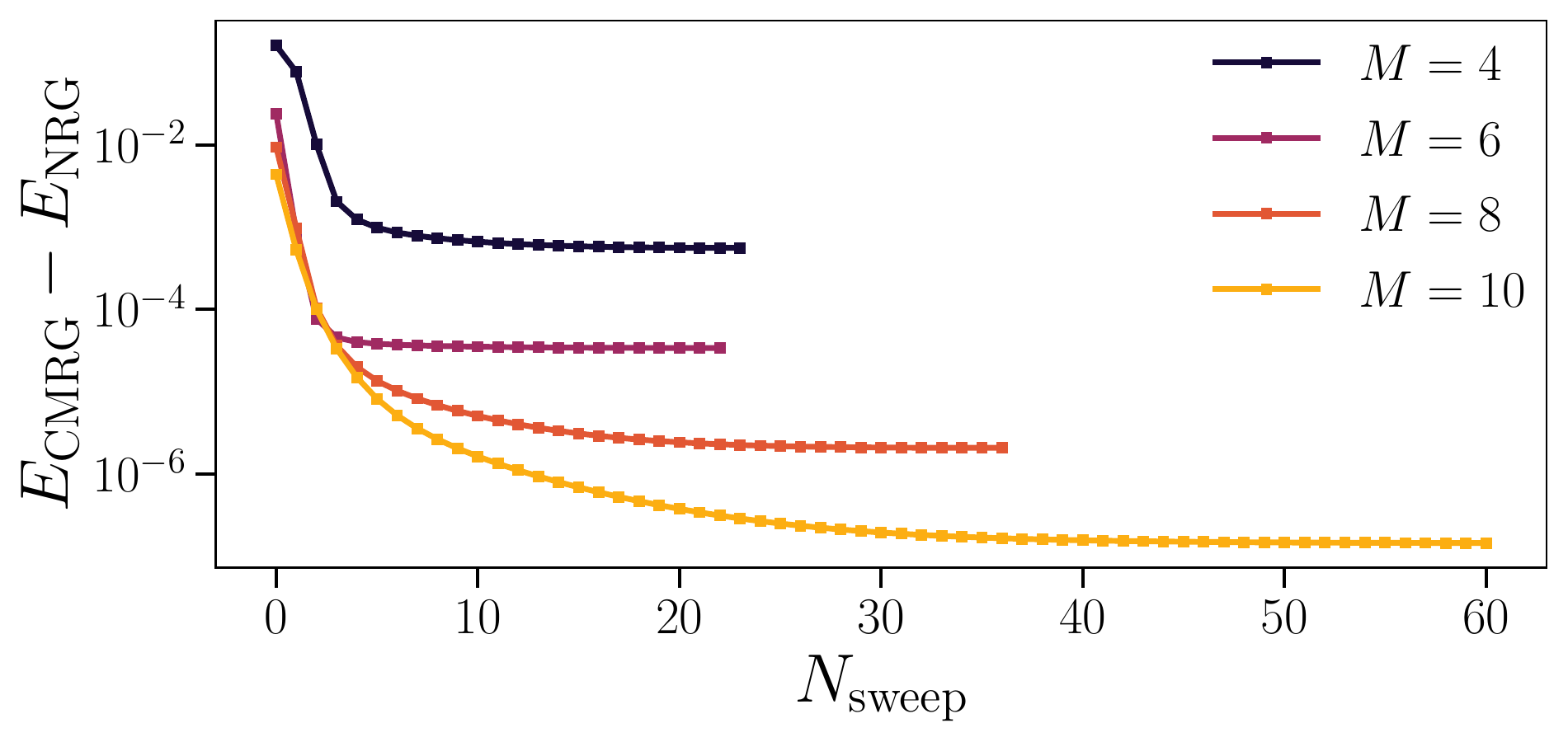}
\caption{Relative ground state energy as a function of the number of sweeps $N_{\mathrm{sweep}}$ for
different numbers $M$ of correlated orbitals. The computation are done with $N=110$ sites on the
Wilson chain and for $U=-0.5D$.}
\label{Speed_Conv}
\end{figure}

Finally, we show on Fig. \ref{Speed_Conv} how the energy converges as a function of sweep iterations,
for several values of the number $M$ of correlated orbitals used in the 
 recursive algorithm. After a rapid
exponential decrease, the energy saturates to a plateau. The sweeps are typically stopped
when the changes in the energy become smaller than $10^{-10}$. We observe that more sweeps are
necessary for larger values of $M$, since more degrees of freedom in the correlated sector need
to be updated. In the shown example, Kondo correlations are well developed, but 
fewer than 10 
sweeps are required to reach an error well below $T_\mathrm{K}\simeq 3.10^{-3}D$.
For $U \ll -D$ or $U > 0$ in the IRLM, correlations are shorter
ranged, and this accelerates reaching the fixed point
to less than $10$ sweeps.

\subsection{Kondo Screening Cloud: clean case}
\subsubsection{One dimensional chain}
We proceed now to the study of a system that goes further than standard NRG calculations on the
Wilson chain, and that exploits the linear scaling of computation time with 
the number of lattice sites for each step within a sweep of the 
 recursive algorithm.
Our goal in this section is to study the quantum correlations in the screening cloud of the IRLM,
a task that requires a good spatial resolution on all scales for exponentially large chains.
The screening cloud of the Kondo model has been studied accurately in early work by
Borda~\cite{Borda_2007}, but the use of NRG is expensive, as each lattice point in the wanted
spatial correlator requires an independent NRG calculation on a two bath geometry. DMRG studies on real
space lattices have also been reported~\cite{Holzner_vonDelft_2009}, but seem limited to lattices of
moderate size. In contrast to NRG, NOs allow for a single shot method, 
that provides a direct computation of the
whole correlation cloud once the many-body wave function is known. In practice, an accurate characterization
of the complete screening cloud can be obtained in a few seconds of running time depending on the size 
of the bath, and this advantage will become crucial for studying ensemble averages later on for the disordered IRLM.

We will still focus on the IRLM Hamiltonian~(\ref{IRLM}), but now using the standard real space
discretization of the finite size system, so that 
the hoppings $t_i=t$ are constant
 on a chain with $N=10^4$ sites.
In order to keep the density of states at the Fermi energy unchanged between the logarithmic and
the real space implementations, we take $t=0.5 D$.
In the spinful Kondo problem, the impurity magnetic moment is screened by the electrons of the bath
below the temperature $T_{\mathrm K}$, and the ground state is a singlet formed by the impurity and the electrons
involved in the screening process. The spatial extension of this singlet, the screening cloud, is usually
observed through the equal-time spin correlator between the impurity and fermions at site $i$,
$\langle \vec{S}.\vec{s}(i)\rangle$.
The one-channel Kondo Hamiltonian and the IRLM are related by a bosonization transformation
\cite{Guinea_Hakim_Muramatsu_1985}, and the $\langle S^z s^z(i)\rangle$ component
of the spin correlator of the Kondo model can readily be related to a natural IRLM observable,
namely the following charge correlator:
\begin{equation}
C^{\pd}_{i} = \langle \big(c^{\dagger}_{1}c^{\pd}_{1} - \langle c^{\dagger}_{1}c^{\pd}_{1} \rangle\big)
\big(c^{\dagger}_{i}c^{\pd}_{i} - \langle c^{\dagger}_{i}c^{\pd}_{i} \rangle \big)\rangle
\label{Ccloud}
\end{equation}
between the impurity charge and the local density operator at site $i$ in the chain.
Substracting here the average charge is done to reveal the long-distance fluctuations that originate
from non-trivial correlations between the impurity and the bath.
At half filling and without disorder, one
has simply $\langle c^{\dagger}_{1}c^{\pd}_{1} \rangle = \langle c^\dag_i c^\pd_i \rangle = 1/2$.
The correlator $C_i$ is computed for every site including the impurity at $i=1$.
By summing $C_i$ over all lattice sites, and using the fact that $\sum_i c^\dag_i c^\pd_i$ is 
the conserved number of particles, one obtains the sum rule:
\begin{equation}
\sum\limits_{i=1}^{N}C^{\pd}_{i} = 0.
\label{sum_rule}
\end{equation}
Without a simple benchmark to NRG results for the observable $C_i$, we compute the variance
of the Hamiltonian $(\langle H^{2}\rangle - \langle H \rangle^{2})/
\langle H \rangle^{2}$ to have a quantitative measure of the precision of the
state built from the truncated NOs. The square root of
the variance is a measure of the error in the variational 
 energy relative to the real ground state energy.
For the results presented below, the states obtained recursively 
give an energy variance of at 
worst $10^{-10}$,
and we will see that other observables are well converged also with respect to $M$.
\begin{figure}[htb!]
\includegraphics[width=1\columnwidth]{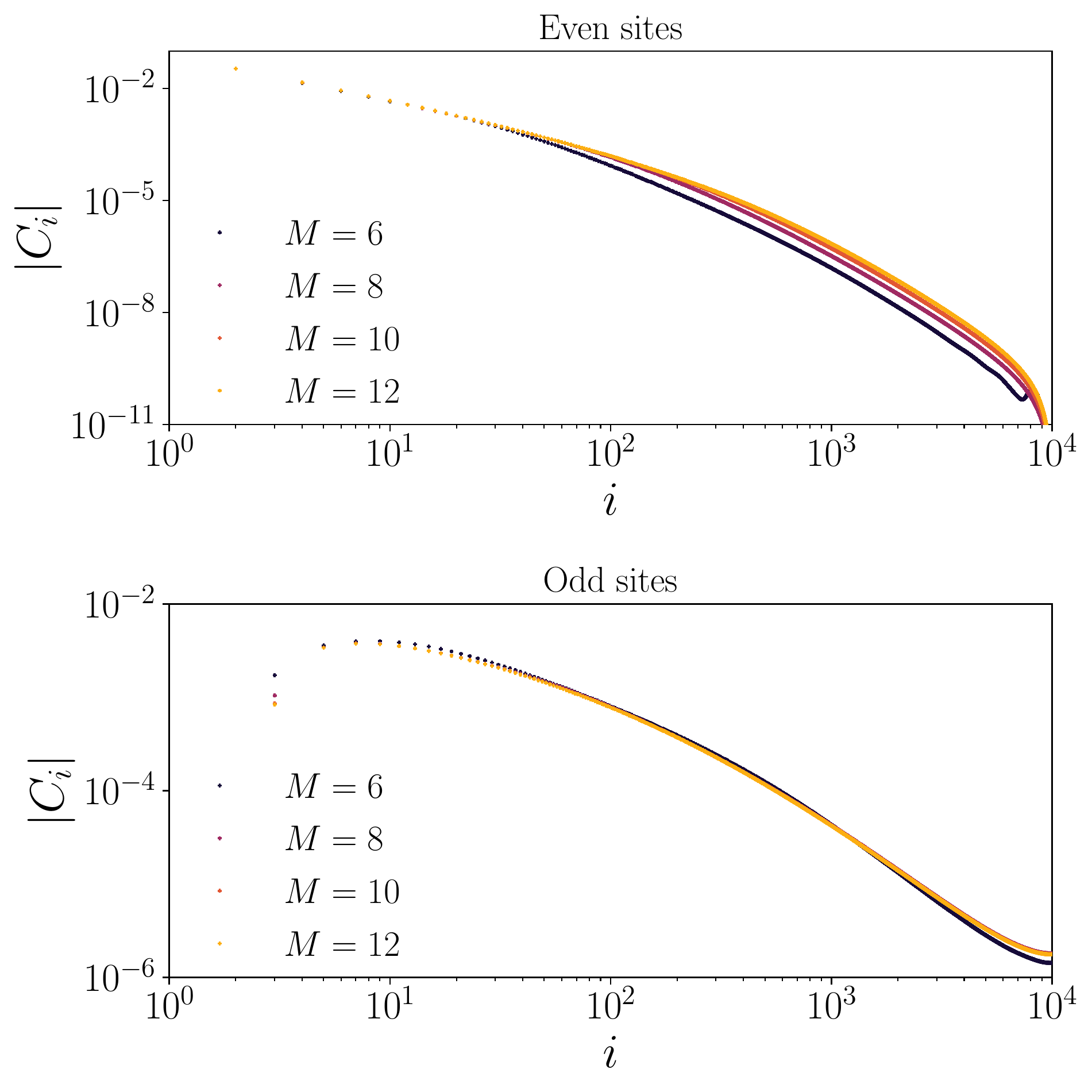}
\caption{Absolute value of the spatial correlator $C^{\protect\pd}_{i}$ at each site of the chain
with $N=10^{4}$ sites (top/bottom panels correspond to even/odd sites). The parameters
are $U=-0.5D$ and $V=0.15D$. Convergence is shown from four increasing values of $M$.}
\label{Cloud_conv}
\end{figure}

Fig. \ref{Cloud_conv} shows our results for the IRLM screening cloud in a very large system with
$N=10^{4}$ sites, for $U=-0.5D$ and $V=0.15D$, corresponding to a screening length $L_{\mathrm{K}} =1/T_\mathrm{K}
\simeq 300$ sites. The even and odd sites have been separated into two panels for better
readability, due to the strong $2k_{\mathrm{F}}$ oscillations of the cloud. This separation allows
us also to extract the enveloppe of these oscillations, which can be compared to
universal scaling predictions.
At intermediate distances $1\ll i\ll L_\mathrm{K}$, perturbation theory~\cite{Barzykin_Affleck_1998}
predicts a decay in $i^{-1}$ for both components. For distances $i\gg L_\mathrm{K}$, an $i^{-2}$ decay
is obtained by Fermi liquid arguments for the largest component, and an $i^{-4}$ decay for the $2k_{\mathrm{F}}$
component~\cite{Ishii_1978}. The full crossover between those two regimes, that has no analytical prediction,
takes place at distances $i \sim L_{\mathrm{K}}$ and extends over a decade.
From these curves, we see that $M=6$ is enough to get a good precision on the screening cloud,
except near the end of the chain, where the $2k_{\mathrm F}$ correlations drop to tiny values. Since 
our simulations are very fast for $M=6$ and $N \sim 10^{3}$ sites (a single run takes a few seconds), we 
can exploit the method to investigate statistical aspects of Kondo correlations in disordered metals, a
challenging question that we will explore in Sec.~\ref{Disorder}.

\subsubsection{Two-dimensional square lattice}

\begin{figure*}[thb!]
\includegraphics[height=.295\textwidth]{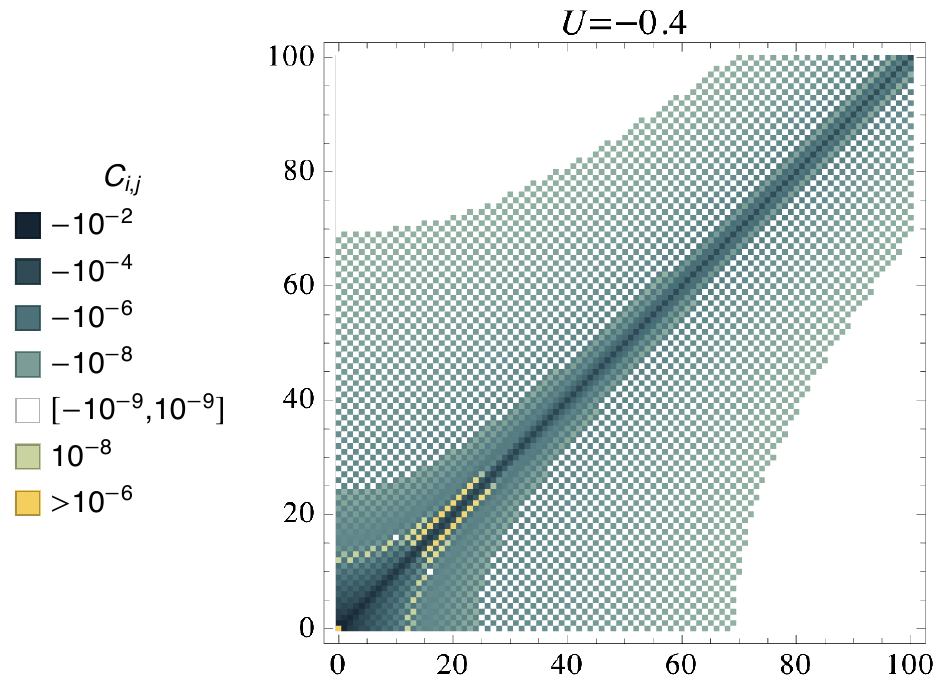}
\includegraphics[height=.295\textwidth]{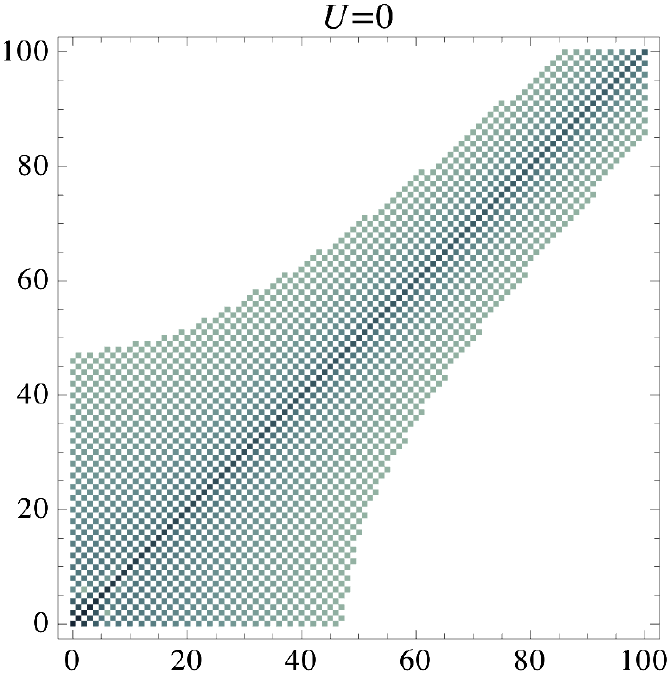}
\includegraphics[height=.295\textwidth]{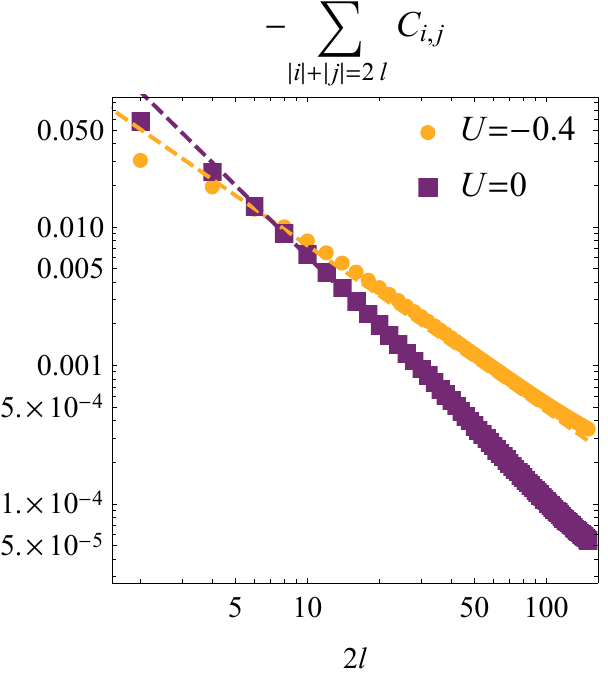}
\caption{\textit{Left panel:} The IRLM screening cloud for a $d$-orbital adatom coupled to the
central site of a $301\times301$ square lattice at half-filling. (One ninth of the full lattice is
shown). Here $V=0.15D$ and $U=-0.4D$ in units of the half-bandwidth $D$. \textit{Middle panel:} Same as 
left panel, but now with $U=0$, showing that correlations are always negative in the non interacting
case. \textit{Right panel:} The screening cloud, summed over all sites that
are $2l$ nearest-neighbor hops away from the central site for $U=-0.4D$ and for $U=0$. The power laws
indicated as dashed lines are $(12l)^{-1.2}$ for $U=-0.4D$ and $(4l)^{-1.7}$ for $U=0$.
\label{2dcloud}}
\end{figure*}

Dynamic impurities embedded in two- or three-dimensional lattices are usually considered under the
assumption of a circular or spherical Fermi surface \cite{Affleck_Review}. 
At length scales sufficiently larger than the 
lattice constant, the approximate circular or spherical symmetry of the problem, together 
with the short range nature of the coupling between the impurity and the host, then permits one
to forget about the lattice and assume s-wave scattering only. This reduces the problem to an effective
one-dimensional model. Until now, the large dimension of the single-particle Hilbert space of
a two- or three- dimensional lattice has prevented the study of higher dimensional screening clouds
beyond the case of circular/spherical Fermi surface. Here we consider an interacting resonant level
adatom coupled to the central site $(0,0)$ of a $(2\Omega+1)\times(2\Omega+1)$ square lattice. 
We assume uniform nearest neighbour hopping in the lattice.
At half-filling the system has a square Fermi surface and a logarithmic van Hove singularity in
the density of states at the Fermi energy. To the best of our knowledge, dynamic impurity screening clouds 
have to date not been studied in this two-dimensional setting.

The Hamiltonian reads:
\begin{eqnarray}
H&=&U(d^\dagger d-1/2)(c_{0,0}^\dagger c^\pd_{0,0}-1/2)+V\left(c_{0,0}^\dagger d+Vd^\dagger c^\pd_{0,0}\right)\nonumber\\
&+&\frac{t}{2}\sum_{\rm NN}
\left(c_{i_1,j_1}^\dagger c^\pd_{i_2,j_2} +c_{i_2,j_2}^\dagger c^\pd_{i_1,j_1}\right),
\end{eqnarray}
where the sum in the last line runs over all distinct pairs of nearest neighbors on a square lattice with opposing corners at $(\pm \Omega, \pm \Omega)$ and at $(\mp \Omega, \pm \Omega)$. We have normalized the hopping term such that the half-bandwidth is $2 |t|$ as in the one-dimensional case we considered previously. 
We study in what follows a $301 \times 301$ lattice.
Point-group symmetries and accidental degeneracies reduce the dimension of the single-particle Hilbert space 
of the non-trivial part of the problem from $90602$ to $11402$, (roughly a factor of $8$ reduction). 
Technical details of the calculation are provided in Appendix \ref{appA}.
In Appendix \ref{appB}, we also refute a claim that a tridiagonalization scheme can reduce
2D Kondo impurity problems on an $L\times L$ lattice to an effective one-dimensional chain with 
order $\mathcal O (L)$ sites. The problem that we tackle maintains therefore a two-dimensional
complexity w.r.t. the orbital sector.

We work in units where the half-bandwidth is $D=1$, 
and we set $V=0.15D$, $U=-0.4D$. This leads to a Kondo temperature
of $T_\mathrm{K}=0.012D$, calculated using (\ref{Tk}). This is somewhat larger than the single-particle 
level spacing $(\pi/302)D=0.010D$ at the Fermi level, which serves as infrared cutoff.
We therefore expect to see well-developed correlations.
Resolving dilute correlations in a two-dimensional lattice requires high accuracy. We
took $M=12$ and estimate that the calculated correlation cloud is converged to correlations down to $\lesssim 10^{-9}$. 
We computed the real space correlation cloud $C_{i,j}=\left<(d^\dagger d-1/2)(c_{i,j}^\dagger
c_{i,j}-1/2)\right>$. In the left panel of Fig.~\ref{2dcloud} we show the result in the patch
$0\leq i,j\leq 100$, which constitutes $1/9^\mathrm{th}$ of the full system studied, using a logarithmic 
color scale clipped at $\pm10^{-9}$.
For comparison, we show the $U=0$ uncorrelated cloud for the same $V$ in the middle panel. At $V=0.15D$, the
$U=0$ cloud has $T_{\rm K}=0.05D$, 5 times larger than $T_{\rm K}$ of the $U=-0.4D$ interacting case. 
In general, we obtain an X-shaped cloud. This is similar to the shape of the single-particle orbital
$b_0$ in the clean lattice at the Fermi energy to which the $d$-orbital couples, which has the 
position-representation wave function
\begin{equation}
\left<0\right| c_{i,j} b_0^\dagger\left|0\right>=\frac{(-)^j\delta_{ij}+\delta_{i0}\delta_{j0}}{2\sqrt{\Omega+1}}.\label{fermiorb}
\end{equation}
However, unlike the Fermi orbital (\ref{fermiorb}), the cloud decays along the $(j,\pm j)$ diagonals and
spreads away from them. It is clear that there is significantly more spreading when $U=-0.4D$, than
when $U=0$, due to the smaller $T_\mathrm{K}$ in the interacting case. A more
qualitative difference is that the $U=0$ cloud never takes on positive values, i.e. the charge polarity
everywhere on the lattice tends to be opposite to that on the $d$-orbital, and is strictly confined to the
$A$-sublattice for which $i+j$ is even. In contrast, for the interacting case with $U=-0.4D$, the charge 
polarity on the central site ($C_{0,0}=0.007$) tends to be the same as on the $d$-orbital, 
as is expected from the fact that negative $U$ favours zero or double occupancy of the $d$-orbital and the 
central site. More interesting is the fact there is another 
``butterfly-like'' region, between 10 and 30 lattice constants away from the central site, where charge polarity again
tends to be the same as on the $d$-orbital (yellow dots in Fig.~\ref{2dcloud}). 
We also observe a significant presence of correlations on the
$B$-sublattice in the interacting case: indeed, at $U=-0.4D$, $\sum_{(i,j)\in B}|C_{i,j}|=0.073$ while 
$\sum_{(i,j)\in A}|C_{i,j}|=0.19$, i.e. correlations on the $B$-sublattice grow to more than a third of the correlations on the $A$-sublattice.
(These $B$-sublattice correlations are strictly zero in the non-interacting case).
Finally, in the right panel of Fig.~\ref{2dcloud}, we plot minus the total amount of correlations at hopping distance 
$2l$, i.e $-\sum_{|i|+|j|=2l} C_{i,j}$, as a function of $2l$. The equivalent quantity in the case of
approximate circular symmetry is $2\pi r C_r$ where $C_r$ is the cloud at any point a distance $r$ from the impurity.
At distances smaller than $1/T_{\rm K}$, $2\pi r C_r$ decays like $1/r$ for all $U$, up to logarithmic corrections. For the case of a square Fermi surface, the sum is dominated by
slowly decaying correlations close to $|i|=|j|=l$. The decay inside the cloud seems to obey a power law, although 
only one decade is clearly resolved. Unlike in the circularly symmetric case, the power law depends on $U$, with 
slower decay at negative $U$ than at $U=0$. Both at $U=0$ and
at $U=-0.4D$, the decay is faster than in the case of circular symmetry. However, the cloud on the diagonals
$(j,\pm j)$ decays more slowly than in the case of circular symmetry. 

The picture of the anisotropic cloud that emerges from this study is one in which significant parts of the region within $1/T_{\rm K}$ away from the impurity is free of significant correlations, while in other regions 
that are $1/T_{\rm K}$ away from the impurity, correlations are asymptotically larger than
expected from the isotropic limit.
Based on these findings, we speculate that in half-filled square lattice with many IRLM type
impurities, a small fraction of impurities will interact via correlations induced in the conduction band, 
possibly even if they are more than $1/T_{\rm K}$ apart, while the majority of the impurities will not 
significantly interact, even if they are closer to each other than $1/T_{\rm K}$.
Extending our study to the case of spinful Kondo impurities would be interesting, but
similar anisotropy effects are certain to take place.

\section{Screening cloud in disordered environments}
\label{Disorder}
\subsection{The model for dirty screening clouds}
Early studies have explored the physics of Kondo impurities in disordered environments, but due
to the difficulties in simulating real space lattices, approximate scaling equations~\cite{Nagaoka_1965}
have mostly been used. Partly 
due to its relevance for dilute Kondo alloys, but partly also due to the above practical limitation,
previous works have focused on the distribution of Kondo temperatures. The recursive NO method promises a
more quantitative description of this
problem, and an access to various microscopic observables in the ground state, such as the screening
cloud. We propose here to add charge disorder to the real space IRLM Eq.~(\ref{IRLM}), still with
uniform hoppings $t_i=t=D/2$ in the chain, adding a generic disorder potential in the Hamiltonian:
\begin{equation}
H_\mathrm{dis}= \sum\limits^{N}_{i=2}v^{\pd}_{i}c^{\dagger}_{i}c^{\pd}_{i}.
\end{equation}
Here the disorder is modelled by a local potential $v_{i}$ that is uniformly distributed in the
interval $[-v,v]$ on every site of the chain, 
except for the impurity site $i=1$.
Indeed, we chose to exclude a random potential on the impurity 
because it automatically
polarizes the impurity site, trivially destroying most 
Kondo correlations.
The potential term $v_i$ explicitely breaks particle-hole symmetry, and the average charge
$\langle c^\dagger_i c^\pd_i\rangle$ is not uniform anymore.
We emphasize that the disordered IRLM corresponds to a Kondo model with random magnetic uniaxial
disorder, a problem that has not been studied extensively yet. The physics that
we explore will turn out to be quite different from known aspects of the standard spinful Kondo problem
with charge disorder.

We use the recursive method presented in Sec.~\ref{RGNO} 
for each realization of the disorder. We take $M=6$ and $N=10^{3}$ sites,
which is enough because charge disorder tends to make correlations shorter ranged than in
the clean IRLM. For a randomly selected subset of disorder realizationss we repeated our
calculations at increasing $M$, and find in each case that indeed the cloud and $T_\mathrm{K}$ data
at $M=6$ is converged. The efficiency of our algorithm 
allows us to sample $10^{4}$ realizations of the disorder,
and obtain good statistics.
For each run, we follow the flow of the ground state energy and of another sensitive
observable, for instance $\langle c^{\dagger}_{1}c^{\pd}_{1}\rangle$, to which the Kondo temperature is related
through Eq.~(\ref{Tk}), so that we can stop the sweeps when both quantities have saturated.
We monitor as well the variance of the Hamiltonian to ensure good convergence
. Out of $10^{4}$ samples, a few tens of disorder realizations failed to converge satisfactorily
within the maximum number of iterations we allowed our program to run. While experimentation
showed that small changes to our sweeping protocol can converge these realizations on 
a case by case basis, we did not pursue it for each case, and rather discarded this
statistically insignificant subset.

\begin{figure}[htb]
\includegraphics[width=1\columnwidth]{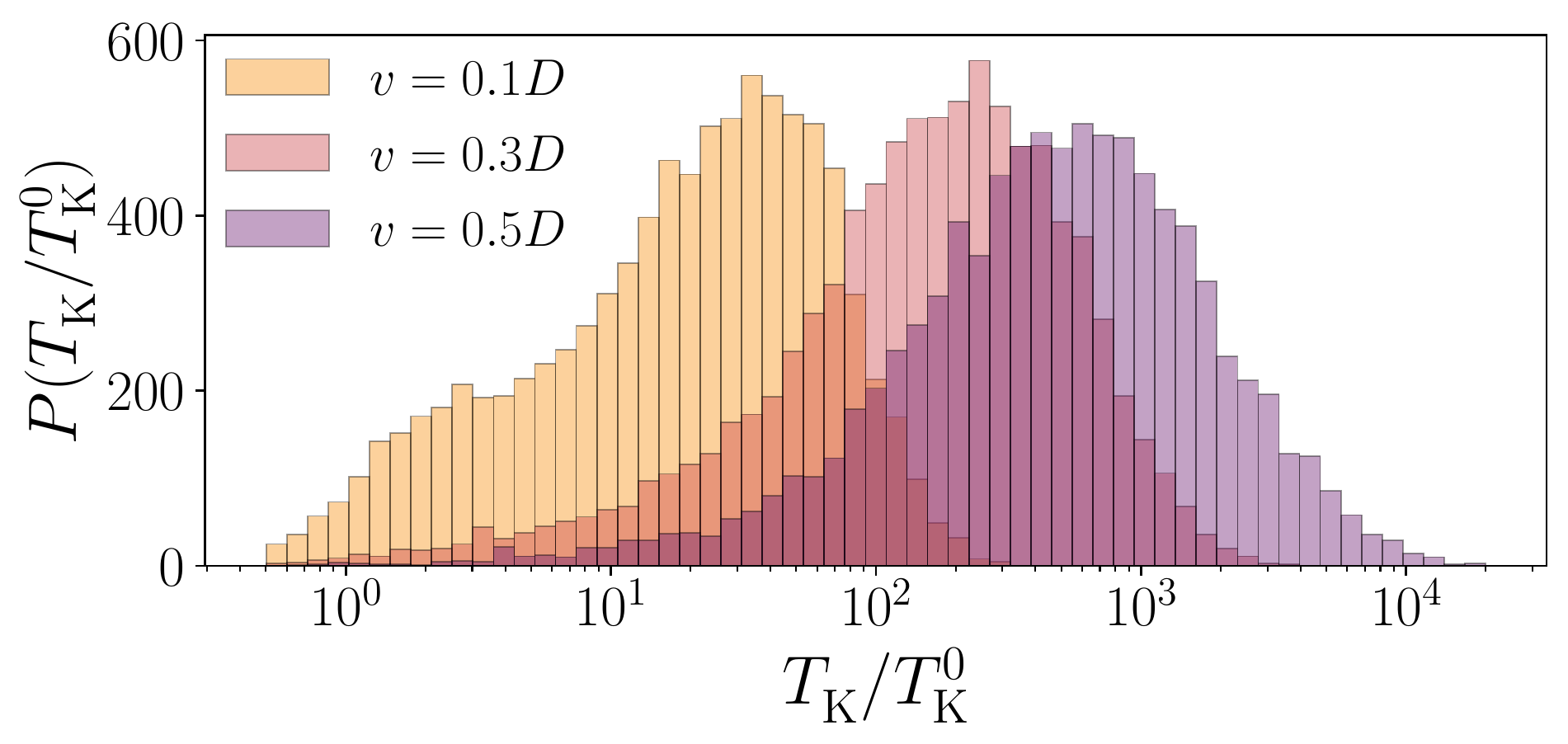}
\caption{Distribution of Kondo temperatures $P(T_{\mathrm K})$ in the Kondo regime of the dirty IRLM
for $U=-0.5D$ and three different disorder strengths. The horizontal axis is normalized to the
clean Kondo temperature $T_{\mathrm K}^{0}\simeq3.10^{-3}D$. Generically, the typical $T_\mathrm{K}$
is pushed to large values, due to a polarization effect on the impurity, which makes Kondo
correlations short range. A clear bimodal distribution is however seen at low disorder, showing
some degree robustness towards values close to the clean value.}
\label{TkDist}
\end{figure}

\subsection{Distribution of Kondo temperatures}
In usual Kondo systems displaying spin correlations and a spin singlet ground state, charge disorder shows
various effects on the distribution $P(T_{\mathrm K})$ of Kondo
temperatures~\cite{Dobrosavljevic_Kirkpatrick_Kotliar_1992,
Kettemann_Mucciolo_2006, Cornaglia_Grempel_Balseiro_2006}.
For weak disorder, the shape of $P(T_{\mathrm K})$ follows a log-normal law that is centered around
the clean Kondo temperature $T_\mathrm{K}^0$.
At increasing disorder, $P(T_{\mathrm K})$ tends to spread away from the clean value $T_\mathrm{K}^0$
with extended tails down to $T_\mathrm{K}=0$. This is because the Kondo scale is proportional to
$e^{-1/(\rho J)}$, with $J$ the Kondo coupling, and charge disorder tends to deplete the local density of states
$\rho$.
A fraction of unscreened moments contribute
to a universal divergence of the magnetic susceptibility with temperature, that is responsible for
an observed non-Fermi liquid behavior as $T \rightarrow 0$ in diluted alloys
\cite{Seaman_Maple_Lee_Ghamaty_Torikachvili_Kang_Liu_Allen_Cox_1991,
Zhuravlev_Zharekeshev_Gorelov_Lichtenstein_Mucciolo_Kettemann_2007,
Moure_Lee_Haas_Bhatt_Kettemann_2018, Miranda_Dobrosavljevic_Kotliar_1997}.
The formation of free magnetic moments appears when disorder opens sufficiently large gaps in the
local density of states at the Kondo impurity, which prevents the formation of the singlet.
As we will see, charge disorder has a drastically different effect in the IRLM, because
Kondo correlations rather develop in the charge sector.

In the case of the disordered IRLM, an important effect of the potential is to drive local charge
offsets of $\langle c^\dag_i c^\pd_i\rangle$ with respect to the clean value 1/2. In addition to random
fluctuations in the density of states, the $d$-level is also Coulomb coupled to site $i=2$, which
induces a strong Hartree shift $U(\langle c^\dag_2 c^\pd_2\rangle-1/2)c^{\dagger}_{1}c^{\pd}_{1}$ that 
polarizes the impurity. This implies that the local charge susceptibility $\chi$ is strongly reduced, and
the relation $T_\mathrm{K}=1/(4\chi)$ leads to a boost of the Kondo energy.
As a result, the distribution $P(T_{\mathrm K})$ shifts towards large values of
$T_{\mathrm K}/T_\mathrm{K}^0$, as is indeed observed in Fig. \ref{TkDist}.
For weak disorder $v=0.1D$, we see an interesting situation where a dominant fraction of the samples are
indeed pushed to large $T_{\mathrm K}$, but a sizeable portion remains weakly affected, as showed by
a bimodal distribution $P(T_{\mathrm K})$. At intermediate disorder $v=0.3D$, long tails towards
$T_{\mathrm K}/T_\mathrm{K}^0\simeq1$ are the remnant of this effect.
Such bimodal distributions are also known to develop in spinful Kondo systems
\cite{Kettemann_Mucciolo_2007, Principi_Vignale_Rossi_2015, Miranda_Dias_da_Silva_Lewenkopf_2014,
Kettemann_Mucciolo_2006}, although the mechanism in the IRLM seems different, as we show now.

\begin{figure}[htb]
\includegraphics[width=1\columnwidth]{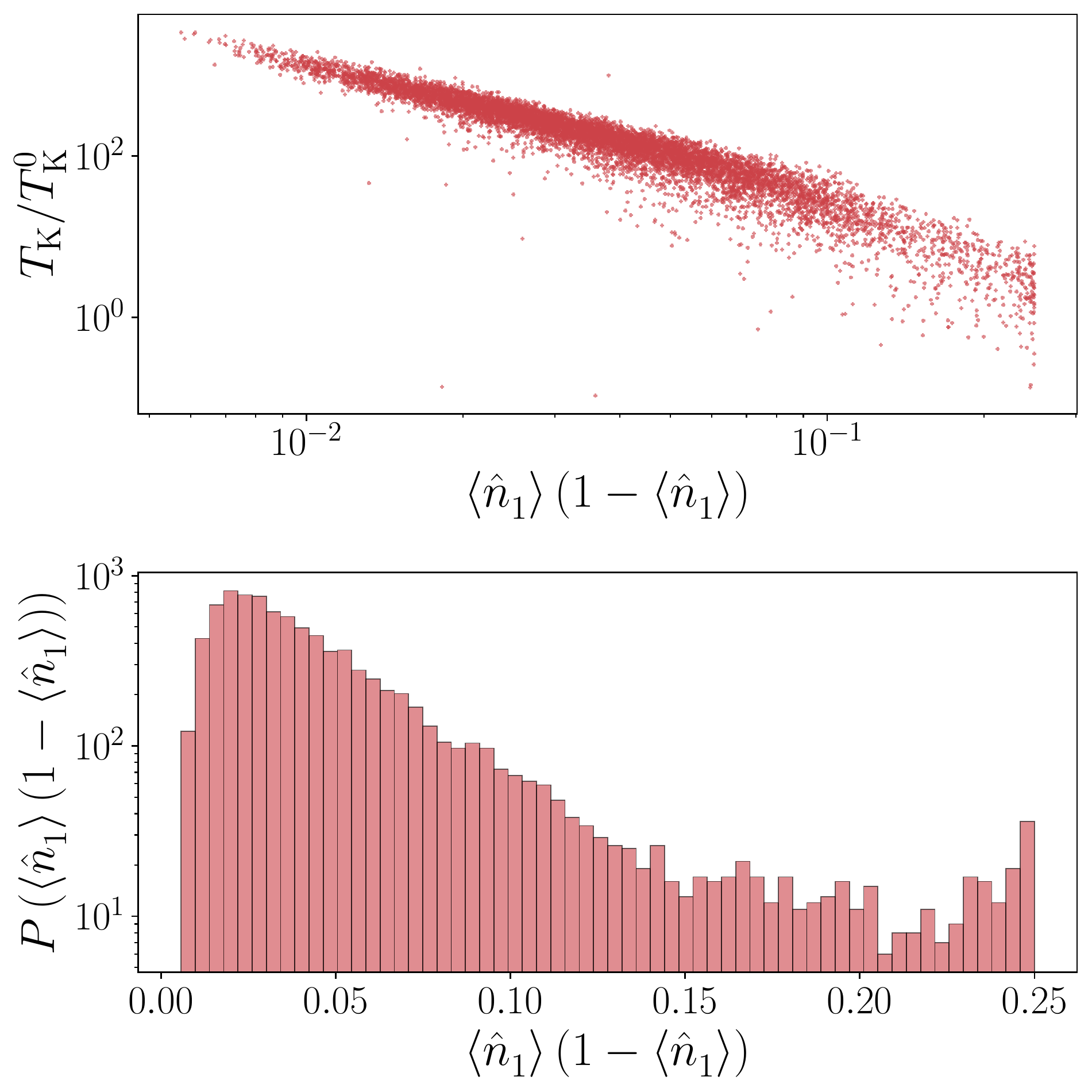}
\caption{\textit{Top panel:} Kondo temperature for $10^4$ samples, plotted as a function of the
``folded'' occupation of the impurity $\langle \hat{n}^{}_{1} \rangle(1-\langle \hat{n}^{}_{1} \rangle)$.
Parameters are the same as in Fig.~\ref{TkDist}, with a strength of disorder $v=0.3D$.
\textit{Bottom panel:} Distribution of the ``folded'' occupation of the $d$-level, showing a
pronounced bimodal structure.}
\label{nd_vs_Tk}
\end{figure}
\subsection{Local charge distribution}
We explore here in more detail the physical mechanism that drives the changes in the Kondo
temperature for the disordered IRLM. Natural orbital methods 
give access to the many-body wave function of each
individual realization, so that we can compute any observable in the ground state.
We advocated that the occupation of the impurity is a key quantity,
and indeed we find a clear statistical correlation between the various Kondo temperatures and the
$d$-level charge (for the same $10^4$ samples as shown previously), see the upper panel
in Fig.~\ref{nd_vs_Tk}. The horizontal axis is here given by
$\langle \hat{n}^{}_{1} \rangle(1-\langle \hat{n}^{}_{1} \rangle)$ instead of
$\langle \hat{n}^{}_{1} \rangle$, which better displays the rare configurations where
$\langle \hat{n}^{}_{1} \rangle$ stays close to 1/2.
The closer the occupation is to $1/2$, which corresponds to a weak breaking of particle-hole symmetry,
the closer the system is to the clean case.
More subtle effects of disorder, such as bimodality, are also exemplified by the distribution
of the $d$-level occupation $P\left(\langle \hat{n}^{}_{1} \rangle(1 - \langle \hat{n}^{}_{1} \rangle )\right)$
that is shown in the bottom panel of Fig.~\ref{nd_vs_Tk}: a large peak is indeed observed around
$\langle \hat{n}^{}_{1} \rangle \simeq 0$ or $1$, and a smaller one around $1/2$, again due to
some rare survivors of the clean state.

\subsection{Study of dirty screening clouds}
We will show that arguments based on the change of Kondo temperature induced by a given disorder
realization give only a
partial view of the internal affairs in quantum impurity ground states. We start here by confirming
microscopically that the rare survival cases where $T_\mathrm{K}$ stays close to the clean value $T_\mathrm{K}^0$
are truly robust to the random potential. This is not totally obvious, since the potential configuration associated
to these cases shows a prominent and fluctuating potential landscape. Recursive generation of
natural orbitals offers an unique way to tackle this problem, by investigating the spatial decay in 
disordered screening clouds, as defined in
Eq.~(\ref{Ccloud}).
\begin{figure}[h!]
\includegraphics[width=1\columnwidth]{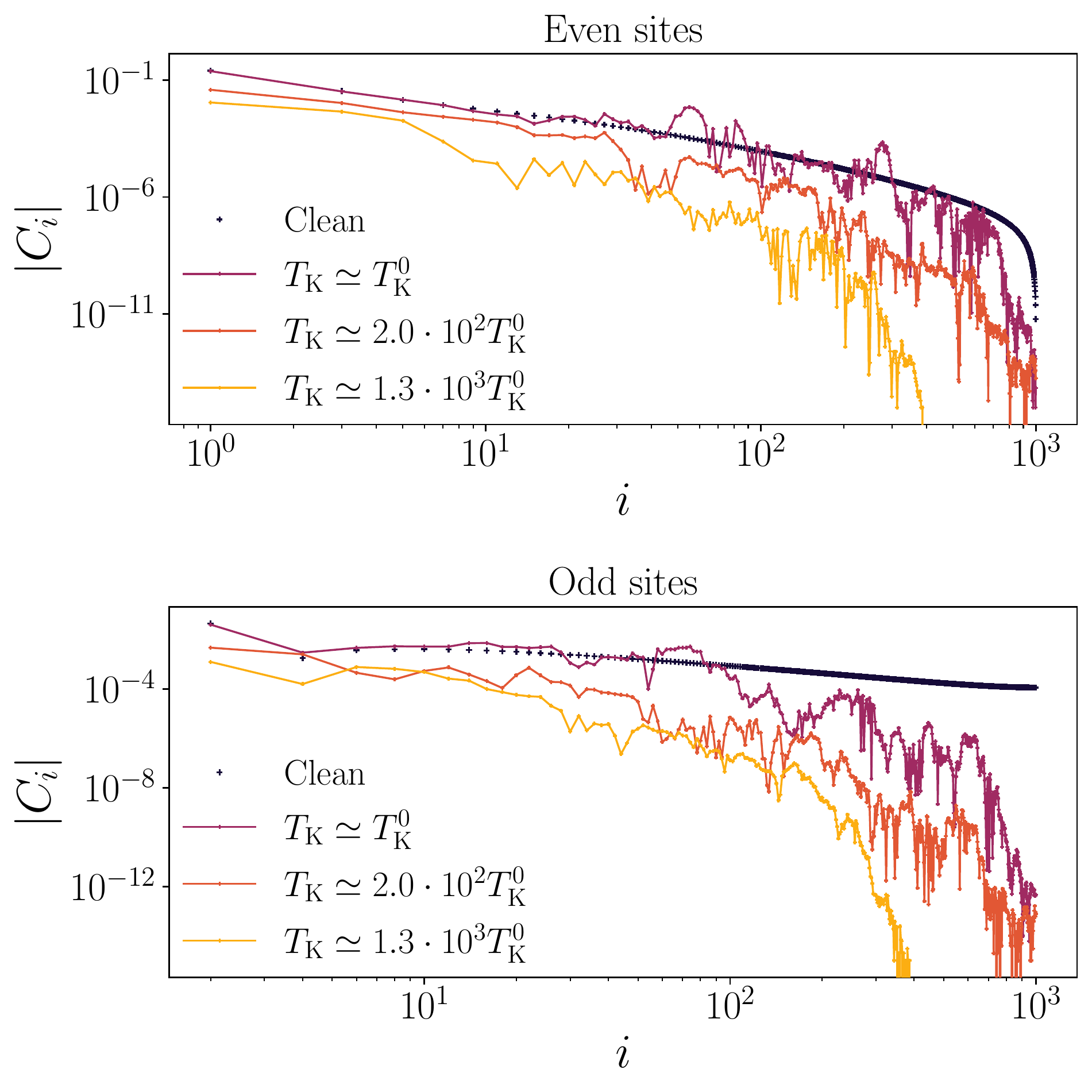}
\caption{Spatial correlator $|C^{\protect\pd}_{i}|$ along the chain for the disordered IRLM at
$U=-0.5D$, $V=0.15D$, and a large disorder amplitude $v=0.3D$. Three specific realizations are
selected with the indicated $T_{\mathrm K}$ values, and the clean screening cloud is shown as 
black dots for comparison
(top/bottom panels correspond to even/odd sites in the chain).}
\label{Cloud_disorder}
\end{figure}
We present in Fig. \ref{Cloud_disorder} three different dirty screening clouds (that are typical of
the thousands of sample that we computed), one with Kondo
temperature $T_{\mathrm K}$ that is close to the clean value, and two that are far from it. The clean
Kondo cloud is also shown as black dots for comparison. We observe clearly that the global amplitude
and shape of the cloud is not radically affected in the case where $T_{\mathrm K}\simeq T_{\mathrm K}^0$
(which corresponds to rare situations),
despite disorder driving large local fluctuations that reflect the underlying profile
of spatially localized orbitals living near the Fermi level, and that couple predominantly to
the impurity. This result is somewhat unexpected because the localization length in our samples
is typically shorter than the clean Kondo screening length. 
In the two cases where $T_{\mathrm K}\gg
T_{\mathrm K}^0$ (which are the most generic), both the amplitude and the spatial extension
$L_{\mathrm K}$ of the cloud reduce, but very surprisingly, $L_{\mathrm K}$ is not decreased in the same
proportion as the observed 100 to 1000 fold enhancement of $T_{\mathrm K}$. Thus, the naive scaling
prediction $T_{\mathrm K}\simeq 1/L_{\mathrm K}$ does not hold for dirty screening clouds, which is 
another unexpected result of our study.
\begin{figure}[htb]
\includegraphics[width=1\columnwidth]{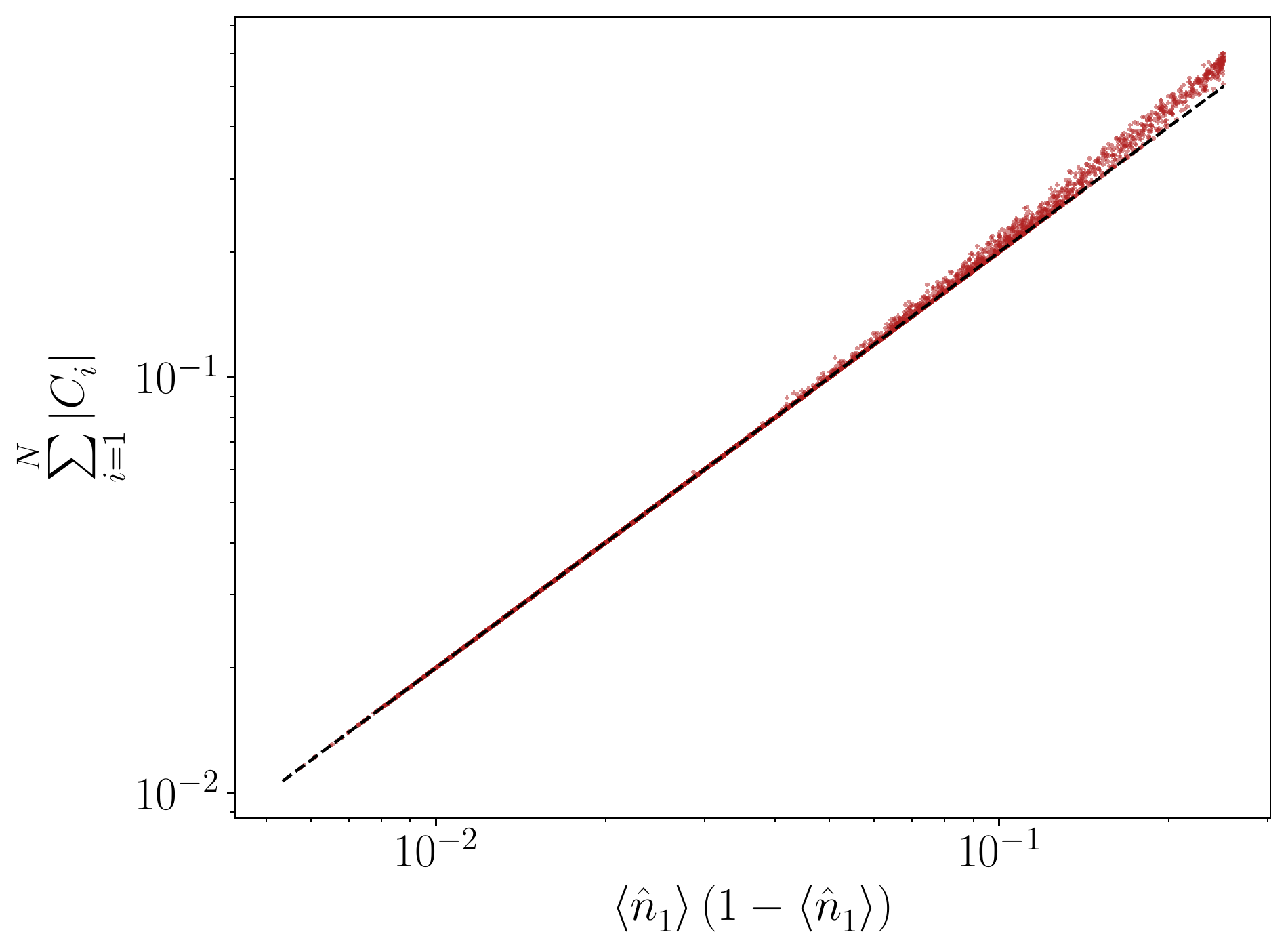}
\caption{Total amplitude of the cloud $||C||$ against the occupation $\langle \hat{n}^{}_{1} \rangle
\left(1- \langle \hat{n}^{}_{1} \rangle \right)$, sampled for all $10^4$ disorder realizations, with the same
parameters as in Figs. \ref{nd_vs_Tk} and \ref{Cloud_disorder}. A nearly perfect match with the
twice the folded occupation, represented by a dashed line is observed.}
\label{Chi_dd}
\end{figure}

We will now show how disorder affects the cloud amplitude, an effect that is again driven by the local
environment of the impurity.
Let us first start by analyzing the Toulouse point, namely $U=0$ in the IRLM. In that case, we can
simplify expression~(\ref{Ccloud}) using Wick's theorem, and we find exactly:
\begin{equation}
C_{i>1} = - | \langle c^\dag_1 c^\pd_i \rangle|^2 <0,
\end{equation}
while $C_1 = \langle \hat{n}^{}_{1} \rangle - \langle \hat{n}^{}_{1} \rangle^2 >0$
since the $d$-level occupancy is positive and smaller than one (this last equation for $C_1$ is
also valid at finite $U$). From the sum rule~(\ref{sum_rule}), we thus get the total cloud amplitude:
\begin{equation}
||C|| = \sum_{i=1}^{N} |C_i| = 2 ( \langle \hat{n}^{}_{1} \rangle - \langle \hat{n}^{}_{1} \rangle^2),
\label{amplitude}
\end{equation}
which is thus only controlled by the impurity occupation.
In the interacting case, Eq.~(\ref{amplitude}) will stay exact provided the correlations
$C_{i>1}$ remain negative, which turns out to be obeyed for most of the $10^4$ samples that were
numerically investigated for the disordered IRLM. We can verify indeed in the Kondo regime at $U=-0.5D$
that the cloud amplitude $||C||$ follows with high accuracy the law of Eq.~(\ref{amplitude})
[See Fig. \ref{Chi_dd}.]
This demonstrates that the local occupation of the dot is a variable that controls very precisely
the global properties of the dirty screening cloud, which may be counter-intuitive at first sight.

\begin{figure}[htb]
\includegraphics[width=1\columnwidth]{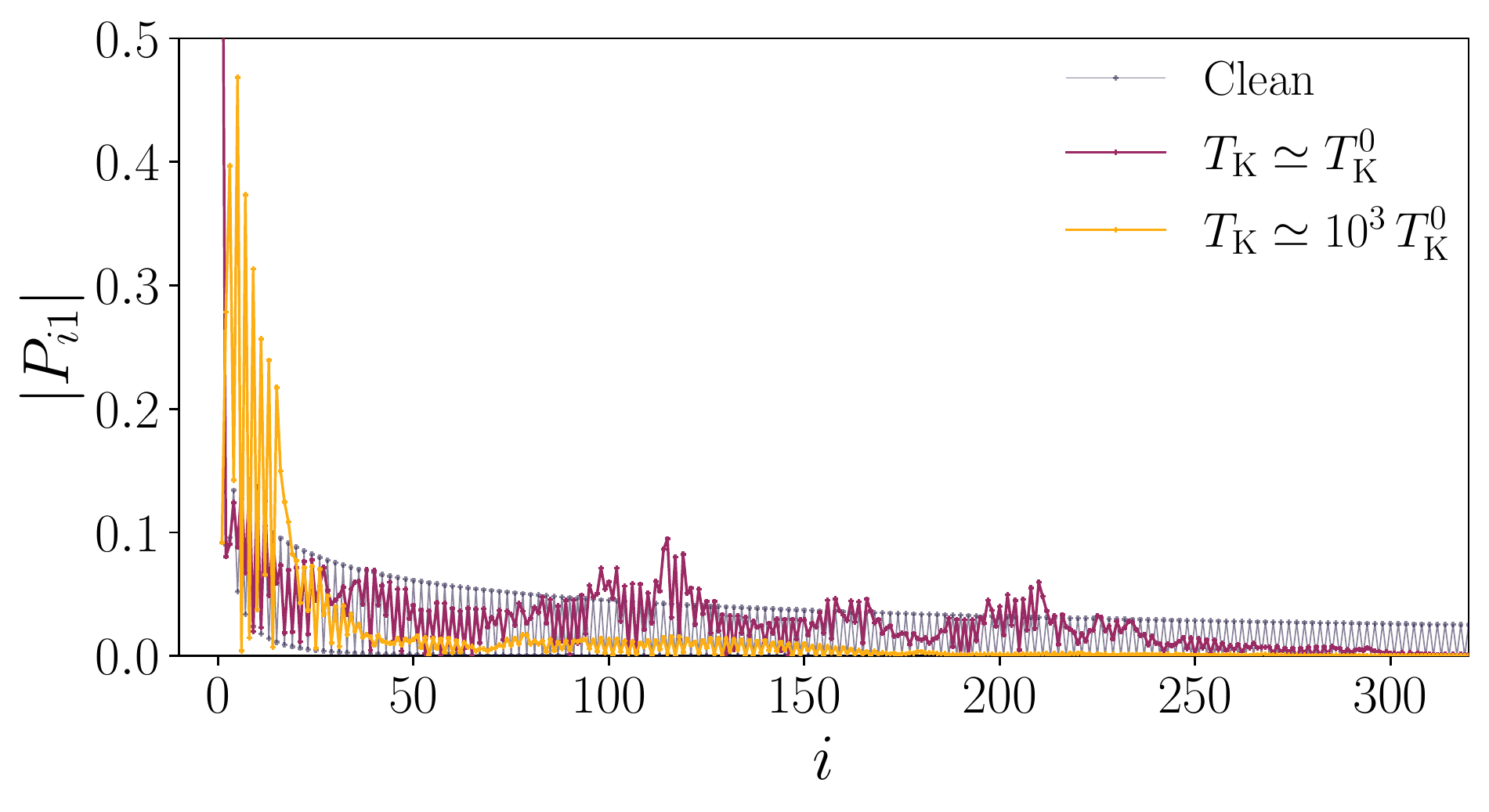}
\caption{Dispersion $|P_{i1}|$ of the most correlated orbital $q^\dagger_1$ along the real-space
chain, with the same parameters as Fig.~\ref{Cloud_disorder}, showing the clean case and two
realizations of disorder, leading respectively to a robust and sensitive Kondo scale. The orbital
is clearly more delocalized for the smallest $T_\mathrm{K}$.}
\label{DirtyOrbitals}
\end{figure}

We can also examine the structure of the Kondo state via the underlying natural orbitals, and
how they react to disorder. Fig.~\ref{DirtyOrbitals} shows the spatial structure of the most
correlated orbital $q^\dagger_1$ given by the squared amplitude $|P_{i1}|^2$.
We compare here the clean case to two dirty
samples, one that is relatively immune to disorder (with $T_\mathrm{K}\simeq T_\mathrm{K}^{0}$),
and one with short-range Kondo correlations (with $T_\mathrm{K}\simeq 10^3 T_\mathrm{K}^{0}$).
As expected, the sample with $T_\mathrm{K}\gg T_\mathrm{K}^{0}$ shows spatial correlations on a
short length scale. This short scale is however significantly larger than $1/T_K$, consistent with our
results for the screening cloud $C_i$.
For the sample with $T_\mathrm{K}\simeq T_\mathrm{K}^{0}$, a strong modulation
of the amplitude is observed with respect to the clean limit, but the structures extends at least up to
the clean Kondo length $1/T_\mathrm{K}$. While confirming again the robustness of some
rare samples to disorder based on the insensitivity of $T_\mr{K}$, we see here that the single 
scale $T_\mathrm{K}$ is insufficient to provide a detailed picture of the spatial Kondo correlations
in dirty metallic hosts.

\section{Conclusion and perspectives}
\label{Conclusion}

We close this article by summarizing our main results and giving perspectives on
the potential applications of recursive natural orbital methods 
to fermionic open quantum systems. 
First, we were able to achieve single-site resolution of the screening cloud in a large 
two-dimensional square lattice at half-filling (with up to 90000 sites).
We showed that rich correlation structures beyond the known s-wave paradigm are manifest,
especially that the spreading is very anisotropic.
Second, we found that spatial correlations between an interacting resonant level and a dirty metallic 
host can be robust to strong charge disorder. We found indeed that the quantum impurity is immune to 
some rare configurations of the random potential, as shown by screening clouds that extend as much in
space as in the clean case, despite strong local fluctuations. However, for the majority of cases where
the impurity is polarized by its environment, the Kondo length of the corresponding cloud is 
less reduced compared to the associated enhancement of the local Kondo scale defined as the inverse 
impurity susceptibility, breaking a scaling hypothesis that applies in the clean case. 
A more visible effect of disorder is a global reduction of the correlation cloud amplitude, that is 
also controlled by local physics at the impurity site.

Our recursive algorithm, that builds on previous ideas on iterative methods for natural orbitals, constitutes
an ideal tool to study the complex electronic environment surrounding a quantum
impurity.
This study prepares the way for 
various extensions, foremost towards a deeper understanding of disordered Kondo
impurities, by applying recursive NO methods to the spinful Anderson impurity model, since natural orbitals provide
an efficient description in that case as
well~\cite{Debertolis_Florens_Snyman_2021,Bravyi_Gosset_2017}.
The protocol we presented here requires one diagonalization of an $N\times N$ matrix per sweep,
while the computation time for all remaining steps scale linearly in $N$, the system size. We are currently 
attempting to developing alternatives to the diagonalization step. This will allow us
to describe large scale electronic baths with up to hundreds of thousand of sites and to include complex 
geometrical effects in 2D or 3D, from interfaces to lithographically designed circuits for quantum
electronics. Also, the entanglement of two (or more) diluted impurities in a metal should receive
some attention, due to the strong anisotropic structure of the screening cloud in 2D and 3D, which
can strongly affect their RKKY coupling.
Besides the study of dirty metallic environments, the interplay of superconductivity
and disorder in quantum impurity physics is also a completely open question~\cite{Zarand_Cloud_SC}.
Recursive natural orbital methods might be able to investigate whether the Kondo singlet between a local 
spin and a dirty superconductor enjoys the protection from disorder associated to Anderson's theorem.
Finally, working with natural orbitals is also the natural language of quantum chemistry, and
an ab-initio implementation for a quantum impurity in metallic hosts would be valuable. Indeed, the
determination from first principles of the right model for a given magnetic atom in a metal is still
an open problem~\cite{Bauerle_Model}, which could be addressed by extensions of natural orbital
methods.

\acknowledgments We thank the National Research Foundation of South Africa (Grant No. 90657).

\appendix

\section{Technical details regarding the IRLM on a square lattice}
\label{appA}

In this appendix, we provide technical details regarding the calculation of the screening cloud
of the IRLM for a $d$-orbital adatom coupled to the central site of a $(2\Omega+1)\times(2\Omega+1)$
square lattice. 
While the single-particle Hilbert space associated with this finite system has dimension 
$(2\Omega+1)^2+1$, the Hamiltonian is invariant under the point group symmetry generated by
$c_{i,j}\to c_{-j,i}$ ($\pi/2$ rotations about the central site) and $c_{i,j}\to c_{j,i}$ (reflection about $i=j$).
As a result, we can consider sectors corresponding to different elements of the point group separately. There are 8 such sectors. The $d$-orbital only couples to the Wannier orbital localized on site $(0,0)$. The latter belongs to the symmetric sector (left invariant by both generators), and as a result
only orbitals in this sector take part in interactions. There are $(\Omega+1)(\Omega+2)/2+1$ orbitals in
this sector. They can for instance be taken as,
\begin{equation}
a_{m,n}=\frac{1}{\sqrt{2}}\sum_{i,j=-\Omega}^\Omega \left[\psi_{m}(i)\psi_{n}(j)+
\psi_{n}(i)\psi_{m}(j)\right]c_{i,j},
\end{equation}
for $m<n$, and
\begin{equation}
a_{m,m}=\sum_{i,j=-\Omega}^\Omega \psi_{m}(i)\psi_{m}(j)c_{i,j},
\end{equation}
together with the $d$-orbital, where
\begin{equation}
\psi_m(j)=\frac{1}{\sqrt{\Omega+1}}\cos\left[\frac{\pi j (m+\tfrac{1}{2})}{\Omega+1}\right].
\end{equation} 
Here $a_{m,n}$ are annihilation operators associated with single-particle states that 
diagonalize the Hamiltonian of the clean system $(V=0)$. The associated single-particle energies
are 
\begin{equation}
E_{mn}=t\left\{\cos\left[\frac{\pi (m+\tfrac{1}{2})}{\Omega+1}\right]+\cos\left[\frac{\pi (n+\tfrac{1}{2})}{\Omega+1}\right]\right\}.
\end{equation}
Accidental degeneracies (not associated with the point group symmetry of the lattice) further reduce the
number of orbitals coupled to the impurity. Let $E_\alpha$ with $\alpha=1,2\ldots,N$, and 
$E_\alpha\not=E_\beta$ when $\alpha\not=\beta$ be the distinct single-particle
energies in the symmetric sector of the clean lattice without the impurity. 
Then, for each $E\in\{E_1,\ldots,E_N\}$ define the normalized superposition
\begin{equation}
b_E=\frac{1}{\sqrt{D_E}}\sum_{m,n} \delta_{E_{m,n},E}\left<0\right|c_{0,0} a_{m,n}^\dagger\left|0\right>a_{m,n}.
\end{equation}
Explicitly $\left<0\right|c_{0,0} a_{m,n}^\dagger\left|0\right>=\sqrt{2-\delta_{m,n}}/(\Omega+1)$
and 
\begin{equation}
D_E=\frac{\sum_{m,n}\delta_{E_{m,n},E}(2-\delta_{mn})}{(\Omega+1)^2}.
\end{equation}
The Wannier orbital
localized on the central site is a superposition of these $b_E$-orbitals:
\begin{equation}
c_{0,0}=\sum_{E=E_1}^{E_N} \sqrt{D_E}\, b_E.
\end{equation}
The $b_E$-orbitals are also eigenstates of the clean lattice tight binding Hamiltonian.
Thus only the $b_E$ orbitals are involved in interactions, and we therefore have to solve a many-body
problem in a Fock space built on $N+1$ orbitals, where $N$ is the number of distinct single-particle energies
associated with the symmetric sector (under point group symmetry) of the clean lattice without the impurity.
The Hamiltonian is of the ``star'' variety and reads
\begin{eqnarray}
H_{\ast}&=&\frac{U}{2}(d^\dagger d-1/2)\left(\sum_{E,E'=E_1}^{E_N} \sqrt{D_E D_{E'}} b_E^\dagger b_{E'}-1/2\right)\nonumber\\ 
&+&V\sum_{E=E_1}^{E_N}\sqrt{D_E}
\left(b_E^\dagger d+d^\dagger b_E\right)
+\sum_{E=E_1}^{E_N} E\, b_E^\dagger b_E.\label{h2deff}
\end{eqnarray}
The energies of the single-particle orbitals that diagonalize $H_{\ast}$ when $U=0$ are roots of the characteristic equation
\begin{equation}
\varepsilon=V^2\sum_{E=E_1}^{E_N}\frac{D_E}{\varepsilon-E}.
\end{equation}
Clearly, there are $N+1$ distinct roots, none of which equal any of the $E_\alpha$. Thus, each $E_\alpha$
is perturbed when the $d$-orbital hybridizes with the lattice, and the number of orbitals coupled to the 
impurity cannot be reduced further. 
Associated to the logarithmic van Hove singularity in the middle of the band is a $\lceil (\Omega+1)/2 \rceil$-fold
degenerate zero-energy-level: $E_{m,\Omega-m}=0$ for $m\leq\Omega/2$. 
Often, this is the only accidental degeneracy that is present, in which case 
\begin{equation}
N=\frac{(\Omega+\Omega\,{\rm mod}\,2)(\Omega+2-\Omega\,{\rm mod}\,2)}{2}+1.
\end{equation}
 
In applying the RGNO algorithm to the above problem, we made two small modifications 
to the protocol presented in Sec.~\ref{RGNO}: Firstly, because single-particle orbitals cannot be ordered
uniquely based on their distance from the central site, and because we are considering a clean system
in which Anderson localization is absent, we pick the initial correlated sector to contain the $M$
$b_E$-orbitals with energies closest to zero (the Fermi level). Secondly, we order orbitals in
the uncorrelated sector by diagonalizing the clean lattice Hamiltonian $\sum_{E=E_1}^{E_N} E\, b_E^\dagger b_E$, projected onto the uncorrelated sector, and ordering eigenstates in ascending
order of their distance from the Fermi energy. Unlike what we did previously, here we do not include
hybridization terms when ordering the uncorrelated sector. We found that this was necessary to obtain
good convergence. We believe this is due to the fact that the large bare hybridization term between the 
$b_0$-orbital and the $d$-orbital
associated with the van Hove singularity at $E=0$ 
($V/ \sqrt{\Omega+1}$ as compared to $\sim V/(\Omega+1)$ at other energies),
is strongly renormalized downward by the interaction.

\section{No exact reduction from $L\times L$ lattices to $\mathcal O(L)$ chains}
\label{appB}

\begin{figure}[htb]
\includegraphics[width=1\columnwidth]{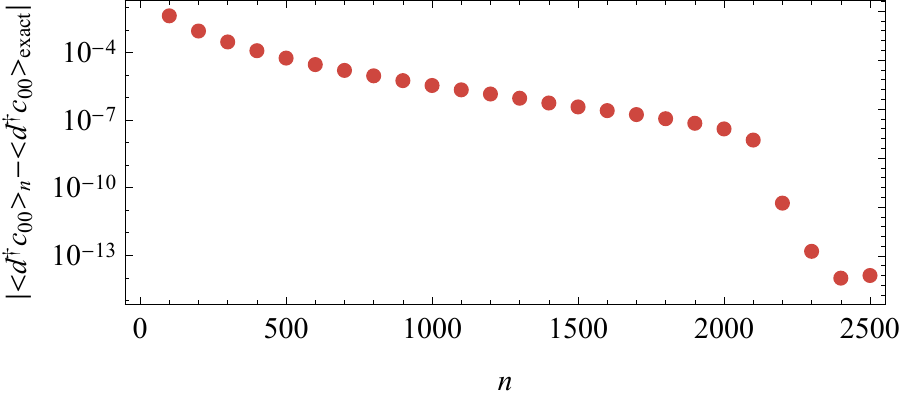}
\includegraphics[width=1\columnwidth]{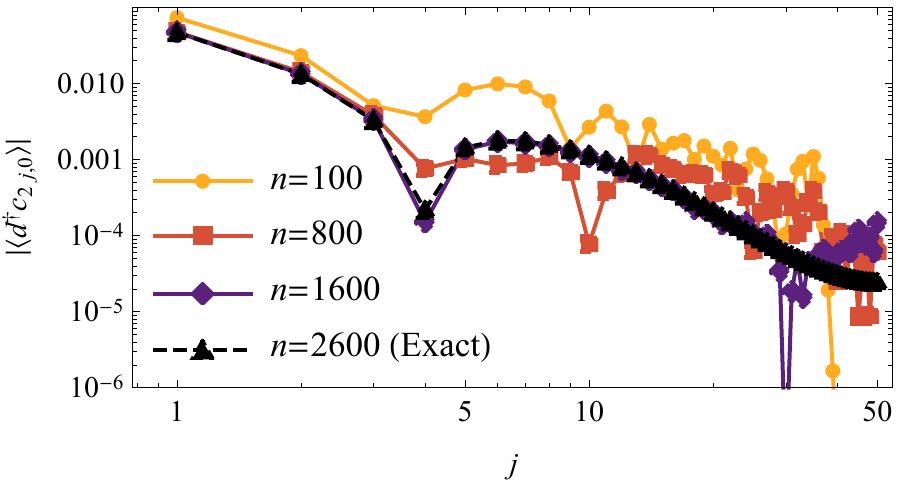}
\caption{\textit{Top panel:} The error in the local hybridization $\left<d^\dagger c_{0,0}\right>$ 
for the non-interacting resonant level model on a diamond shaped lattice with boundaries at $(i,j)$ such 
that $|i|+|j|=100$, after $n$ iterations of the recursive scheme presented in Ref.~\cite{Feiguin_Cloud_2D}. Here
$V=0.15D$ in units of the half-bandwidth $D$.  \textit{Bottom panel:} spatial profile of the non-local 
hybridization $\left<d^\dagger c_{2j,0}\right>$ as a function of site $j$, for the same system as in the 
top panel, after $n$ iterations, for various $n$. These results demonstrate that no sizeable
gain can be obtained using 1D Lanczos orbitals, beyond the symmetry reduction already performed in 
Appendix~\ref{appA}. }
\label{figFeig}
\end{figure}

The preceding appendix shows that symmetry considerations can reduce the numerical effort
on a $L\times L$ lattice from $L^2$ orbitals down to $L^2/8$.
However, in Ref.~\cite{Feiguin_Cloud_2D} a scheme is presented that maps clean nearest neighbor 
tight-binding Hamiltonians in higher than one dimension onto one-dimensional chains with nearest 
neighbor hopping only. The scheme works by recursively generating orthogonal ``Lanczos'' 
orbitals $\left|\psi_n\right>$ on the higher dimensional lattice, such that
$\left<\psi_{m}\right|H\left|\psi_{n}\right>=0$ for $m<n-1$, where $H$ is the higher-dimensional tight-binding
Hamiltonian. While this method is numerically exact once all symmetry class orbitals have been
exhausted, it is claimed~\cite{Feiguin_Cloud_2D} that one can keep in an exact way only $\mathcal O(L)$ 
sites on the effective chain, much lower than the expected $\mathcal O(L^2)$ effort.
In view of the analysis presented in Appendix \ref{appA}, this claim cannot be correct. The point group 
symmetry of the square lattice only leads to a reduction from $L^2$ to $\sim L^2/8$ before one ends up with 
a system with a non-degenerate single-particle spectrum in which each level is coupled to the impurity.

However, it is tempting to ask whether an early truncation of the chain made before reaching its final 
end could lead to an accurate approximation of the impurity problem.
We therefore implemented the scheme of Ref.~\cite{Feiguin_Cloud_2D}
and studied the non-interacting resonant level model ($U=0$) for a $d$-orbital coupled to the
central site of a diamond-shaped lattice with corners at $(\pm \Omega,0)$ and $(0,\pm \Omega)$. 
(This is the shape recommended in Ref.\,\cite{Feiguin_Cloud_2D}.) We
took $V=0.15$ and $\Omega=100$. The total number of sites, excluding the $d$-orbital is $20201$. We
included the stabilization measures discussed in Ref.~\cite{Feiguin_Cloud_2D} and find that the method is 
indeed numerically stable. If the 
claims in Ref.~\cite{Feiguin_Cloud_2D} were correct, we would have obtained the exact answer after 
$100$ iterations.
In the top panel of Fig.~\ref{figFeig} we plot the error in the hybridization 
$\left<d^\dagger c_{0,0}\right>$ as a function of
the number $n$ of iterations. After $100$ iterations, $\left<d^\dagger c_{0,0}\right>$ is obtained
only to two significant digits. After $2000$ iterations the answer is correct to seven digits, and
convergence to the exact answer (to numerical precision) is obtained finally after the
expected $L^2/8\simeq2600$ iterations.
In the bottom panel of Fig.~\ref{figFeig}, we plot the spatial profile of the hybridization
from the impurity to the bath along the $x$-axis, $|\left<d^\dagger c_{2j,0}\right>|$ after respectively 
$100$, $800$, $1600$ and $2600$ iterations. 
We see that more and more Lanczos iterations are required to get reasonable accuracy, the further away from
the central site one moves.  We conclude that the recursive scheme of Ref.~\cite{Feiguin_Cloud_2D} 
does not offer a shortcut for accurately calculating the impurity screening cloud on large 2D lattices.

\bibliography{RGNO_biblio.bib}

\end{document}